\documentclass[aps,prstab,superscriptaddress,twocolumn]{revtex4-1} 
\usepackage{lineno}
\usepackage[utf8]{inputenc} 

\usepackage{booktabs} \usepackage{array} \usepackage{verbatim} \usepackage{cprotect}

\usepackage{graphicx} \usepackage{grffile}
\graphicspath{{./figs/}}
\usepackage{subcaption}
\usepackage{amsmath}

\usepackage[super,negative]{nth}
\usepackage[section]{placeins} \usepackage[tight-spacing=true]{siunitx}

\usepackage[hidelinks]{hyperref}
\usepackage{tikz}
\usetikzlibrary{shapes,arrows,graphs}

\usepackage{gensymb}

\begin{document}

		\title{Medical therapy and imaging fixed-field alternating-gradient accelerator with realistic magnets}

\affiliation{The University of Manchester, Oxford Road, Manchester M13 9PL, UK}
\affiliation{ASTeC, STFC Daresbury Laboratory, Daresbury, WA4 4AD, UK}

\author{S. Tygier}
\email{sam.tygier@manchester.ac.uk}
\affiliation{The University of Manchester, Oxford Road, Manchester M13 9PL, UK}
\affiliation{The Cockcroft Institute, Daresbury, WA4 4AD, UK}

\author{K. Marinov}
\email{kiril.marinov@stfc.ac.uk}
\affiliation{ASTeC, STFC Daresbury Laboratory, Daresbury, WA4 4AD, UK}
\affiliation{The Cockcroft Institute, Daresbury, WA4 4AD, UK}

\author{R.B. Appleby}
\email{robert.appleby@manchester.ac.uk}
\affiliation{The University of Manchester, Oxford Road, Manchester M13 9PL, UK}
\affiliation{The Cockcroft Institute, Daresbury, WA4 4AD, UK}

\author{J. Clarke}
\affiliation{ASTeC, STFC Daresbury Laboratory, Daresbury, WA4 4AD, UK}
\affiliation{The Cockcroft Institute, Daresbury, WA4 4AD, UK}

\author{J.M. Garland}
\affiliation{The University of Manchester, Oxford Road, Manchester M13 9PL, UK}
\affiliation{The Cockcroft Institute, Daresbury, WA4 4AD, UK}

\author{H. Owen}
\affiliation{The University of Manchester, Oxford Road, Manchester M13 9PL, UK}
\affiliation{The Cockcroft Institute, Daresbury, WA4 4AD, UK}

\author{B. Shepherd}
\affiliation{ASTeC, STFC Daresbury Laboratory, Daresbury, WA4 4AD, UK}
\affiliation{The Cockcroft Institute, Daresbury, WA4 4AD, UK}

\begin{abstract}
NORMA is a design for a normal-conducting race track fixed-field alternating-gradient accelerator (FFAG) for protons from 50 to 350~MeV. In this article we show the development from an idealised lattice to a design implemented with field maps from rigorous two-dimensional (2D) and three-dimensional (3D) FEM magnet modelling. We show that whilst the fields from a 2D model may reproduce the idealised field to a close approximation, adjustments must be made to the lattice to account for differences brought about by the 3D model and fringe fields and full 3D models. Implementing these lattice corrections we recover the required properties of small tune shift with energy and a sufficiently-large dynamic aperture. The main result is an iterative design method to produce the first realistic design for a proton therapy accelerator that can rapidly deliver protons for both treatment and for imaging at up to 350~MeV. The first iteration is performed explicitly and described in detail in the text.
\end{abstract}

\maketitle

\section{Introduction}
\label{sec:introduction}

\subsection{Proton therapy and the need for imaging}

Modern radiotherapy is a mainstay of cancer treatment today, and in the developed world around half of cancer patients will receive radiotherapy as part of their treatment. The bulk of radiotherapy treatments are given by using x-rays from a low-energy linac (c. 10-20~MeV kinetic energy), where the method of intensity-modulated radiotherapy (IMRT) utilizes numerous treatment field directions and a moveable set of multi-leaf collimators that may obtain a highly-conformal dose to a prescribed treatment volume~\cite{Webb:1997ui,Nutting:2000vy,Webb:2001wu,Garden:2004ht,Bortfeld:2006jd,Veldeman:2008gt,Staffurth:2010dh}. Whilst accurate pre-treatment patient imaging (such as computed tomography) are of course crucial to creating and delivering the planned treatment, the inherent near-exponential reduction of x-ray intensity with depth makes x-ray treatment comparatively less sensitive to errors in the patient density that is inferred from the patient imaging process.

Proton therapy is an alternative method of delivering a radiotherapeutic treatment to a patient, already known since 1947~\cite{wilson47} to potentially offer an inherently more precise delivered dose since the proton stopping transfers those particles' kinetic energy to deposited dose according to the Bethe-Bloch equation~\cite{owencontempphys}. A well-known initial proton energy coupled with a well-known tissue density allows one to place the maximum dose - at the Bragg peak at the end of the particle range - at a desired position within the patient. The overlap of numerous Bragg peaks from protons with differing initial energy allows a conformal dose in the treatment volume whilst potentially better sparing the surrounding tissue from unwanted dose, particularly important for nearby organs at risk. That said, proton therapy is not thought to be advantageous over x-ray therapy for all radiotherapy treatments, and is often prescribed for particular complex or pediatric treatments where the unwanted ancillary dose may cause a later induction of secondary cancers as a treatment side-effect.

The treatment advantages of proton therapy have led to the creation of the more than fifty operating centers around the world today~\cite{durantecharged2010,Ma:2012um}. Supplanting early facilities at research laboratories, today's hospital-based centers predominantly utilize cyclotrons although synchrotrons are also used. Modern cyclotrons - particularly superconducting ones - offer a number of advantages in terms of simplicity, capital cost and possible dose rate at the patient; 1~Gy may be accurately delivered to a patient by such a source in less than a minute using an average current of $<$1~nA~\cite{owencontempphys}. However, higher-intensity cyclotrons are typically limited to around 230-250~MeV kinetic energy due to relativistic effects, and their fixed-energy extraction requires the use of a mechanical degrader (typically graphite) to lower the proton energy for shallower proton dose delivery. 230~MeV protons are sufficient for treating adult patients (this energy corresponds to about a 33~cm proton range in water), but some treatments would benefit from more rapid (and therefore finer) variation of the energy than is readily achieved using cyclotrons. The latter benefit has led to the proposal both of linacs and of FFAGs (fixed-field alternating-gradient accelerators) for particle therapy; each offers the possibility of obtaining a rapid change of the delivered proton energy at rates up to 1~kHz. Various FFAG designs have therefore been put forward, a notable example being the French RACCAM study that showed how such an accelerator could operate with multiple treatment rooms~\cite{Antoine2009293}.

Irrespective of the accelerator technology used to deliver the particles, the benefits of proton therapy are today limited by inadequacies in the patient imaging used to estimate the patient tissue density~\cite{owencontempphys}. This determination is crucial to deriving a suitable treatment plan, and is further complicated by the complex scattering effects around inhomogeneities that demands Monte-Carlo dose estimation~\cite{Paganetti:2012ct}. X-ray computed tomography is seen as somewhat inadequate as the patient density and composition derived from the Hounsfield measurement can lead to several millimeters of error in the resulting proton range~\cite{Schneider:1999kr,Lomax:2008gw,Knopf:2013bm}. Proton tomography is a better, more direct, measure of the desired proton stopping power and therefore several groups are developing clinical proton tomography instruments that typically track millions of individual protons to assemble a three-dimensional image with a resolution approaching a millimeter~\cite{coutrakonscanner,allinsonpct2015}. But, proton tomography requires protons of sufficient energy to pass through the part of the patient to be imaged, implying significantly-higher incident energies than those that would be used to deliver a (stopped beam) treatment in that same volume. A proton source of 250~MeV could be used for imaging through smaller thicknesses, but patients requiring treatment with 230~MeV protons of course require imaging with much higher proton energies - perhaps as high as 330~MeV or more depending on the image resolution required. No commercial cyclotron today offers this higher energy, and whilst linacs and synchrotrons both in principle could offer such an energy only ProTom has offered a 330~MeV system commercially~\cite{protom}. This lack has previously motivated both the PAMELA design study~\cite{peach2013pamela} - which examined a combined accelerator system offering both protons and carbon ions - and our NORMA study~\cite{garland2015normal,tygier2015pyzgoubi}. Both designs aimed at providing proton energies suitable for tomography, but in the latter proton-only NORMA study the additional goal has been to offer a simple, robust design.

\subsection{FFAGs for proton therapy and imaging}

In our previous work on NORMA we concluded that an FFAG design utilizing an FDF (focusing-defocusing-focusing) arrangement of normal-conducting gradient dipoles was the best way to achieve our desired 350~MeV. The maximum field of around 1.6~T necessitates a somewhat larger circumference than could be obtained with superconducting magnets, but allows for an easier control of the tune during proton bunch acceleration using the higher-order field components in the magnets. Injection of a single proton bunch would be obtained from a cyclotron at an energy of at least 50~MeV, and both injection and extraction would be via a conventional pulsed kicker/septum combination that may benefit from lengthening two of the straight sections between the 10 FDF cells to give a racetrack layout; the acceleration cycle of around 1~ms per extracted single bunch would enable rapid bunch-by-bunch variation of the delivered energy at the patient. A more detailed discussion of the accelerator magnet lattice design is given in Garland et al.~\cite{garland2015normal}, and is summarized in section~\ref{sec:norma_lattice}. In this previous article both the round and racetrack configurations of NORMA are considered; however, here we focus on just the round variant.

The optics design of NORMA presented in~\cite{garland2015normal} was performed with idealized magnet models, to enable the overall beam dynamics to be studied and optimized. This is a common approach that allows optimization using a reduced number of variables, and gives a tractable simulation time. Our idealized magnet models use analytic expressions for the magnetic field within the body of the magnet, and a simple analytic expression for the fringe field fall-off; this is a compromise between simplicity and realism. However, once a design has been obtained with idealized magnets it is of course important to check that the approximations used do not have a significant effect on the dynamics of the accelerator. For example, the presence, size and shape of the realistic fringe fields will affect the focusing of a magnet and the fields from the physical magnets will not completely match analytic models. Their shape can introduce higher-order effects that are not expressed in their ideal analytic form and manufacturing tolerances will cause deviations from the ideal field. This is an issue that must in particular be addressed in FFAG design - where the magnets are large in aperture and inherently nonlinear in nature - before one can be confident that a realistic and therefore buildable design has been obtained.

In this article we build on our previous work~\cite{garland2015normal}, and discuss the detailed 2D and 3D magnet modelling carried out to improve the realism of the NORMA design. Realistic magnet fields introduce perturbations on the idealized dynamics such as tune shift and a reduction in the dynamic aperture. We show how this can be mitigated - by re-matching and re-optimizing - in order to recover acceptable dynamical properties; we demonstrate the recovery of a sufficiently flat tune profile and sufficient dynamic aperture for injection and acceleration. We introduce the magnet models in three stages, so as to methodically understand the importance of different effects.

The layout of this paper is as follows. In section~\ref{sec:norma_lattice} we describe the original NORMA lattice and the methods used for modelling complex elements and performing long-term stability studies. This is followed in section~\ref{sec:magnet_design_simulations} by the magnet design process. In section~\ref{sec:2d_magnets} NORMA is modelled using the radial profile from the 2D magnet simulations, with realistic transverse fields but with a simplified fringe model; the effects on the dynamical properties are shown to be small. Then in section~\ref{sec:fringe_fields} we show the effects of changing the parameters used for the fringe fields, and how these can be mitigated by rematching the overall lattice; the results show that all realistic likely fringe field perturbations to the dynamical properties can be absorbed by retuning the strength and field index of the magnets. Finally, in section~\ref{sec:3d_magnets} we model NORMA with a full set of 3D magnet models. Here the effects on the dynamics are more significant, and so we show the necessary corrections to the field profile to restore the dynamical properties and crucially of the dynamic aperture.
The overall result is a method for the iterative design of realistic magnets for a medical FFAG capable of delivering 350~MeV protons for imaging, and a demonstration of the first step of the iteration.

\section{The NORMA Accelerator}
\label{sec:norma_lattice}

In this section we introduce the nominal NORMA design, as described in detail in~\cite{garland2015normal}, as well as the code Zgoubi~\cite{zgoubi} which is used for tracking particles through the lattice.

\subsection{The NORMA lattice}

The round NORMA lattice design variant is a FFAG consisting of 10 identical FDF triplets. It accelerates proton bunches to 350~MeV, with injection from a cyclotron with at least 50~MeV kinetic energy. NORMA utilizes normal-conducting sector magnets with a scaling FFAG field achieved by pole-face shaping. Within the 36\degree~cell, each magnet has a sector angle of 6\degree; within the FDF triplet the magnets are spaced by 1.8\degree. The scaling FFAG magnets result in a flat tune (tune shift below \SI{e-3}) over the full energy range from 50 to 350~MeV. The magnet field fall-off is modelled using Enge-like fringe fields with a 6~cm extent~\cite{enge1964effects,muratori2015analytical}.

\begin{figure}[htbp]
  \centering
  \includegraphics[width=0.9\columnwidth]{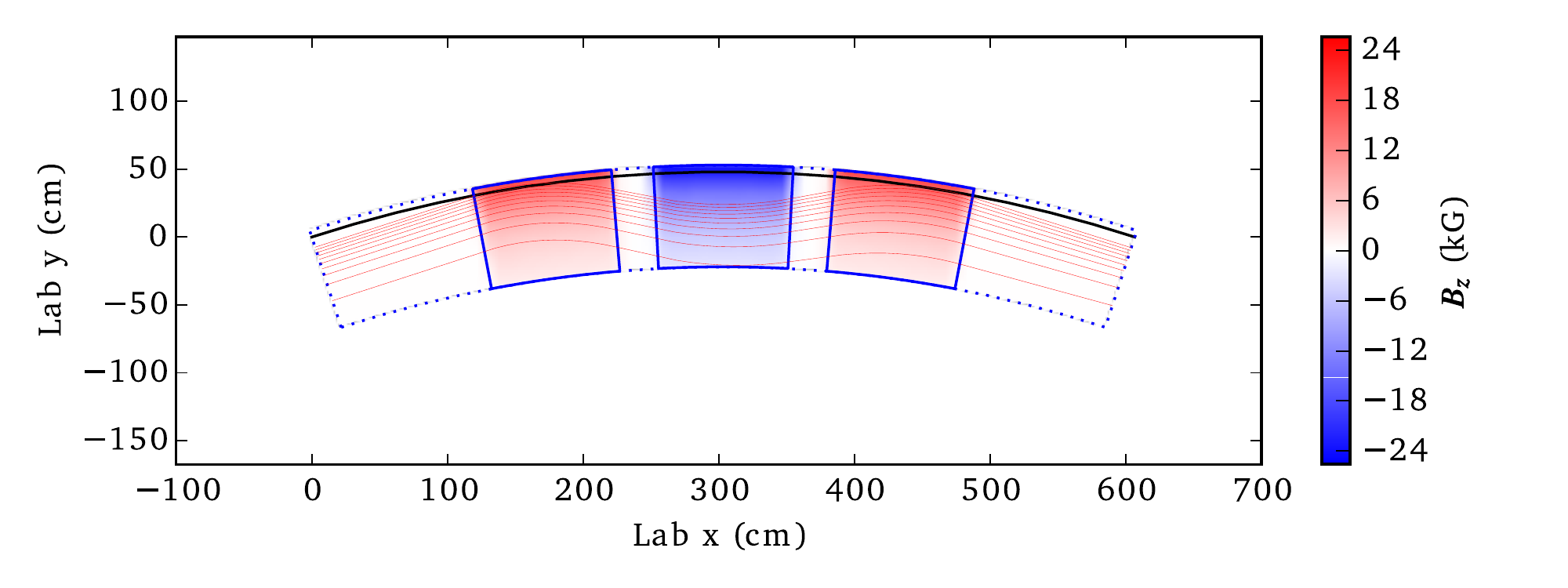}
  \caption
{NORMA cell, showing the mid-plane magnetic field strength in the three scaling FFAG magnets forming the FDF triplet, and the closed orbits for a range of energies.}
\label{fig:norma_cell}
\end{figure}

\begin{figure}[htbp]
  \centering
  \includegraphics[width=0.9\columnwidth]{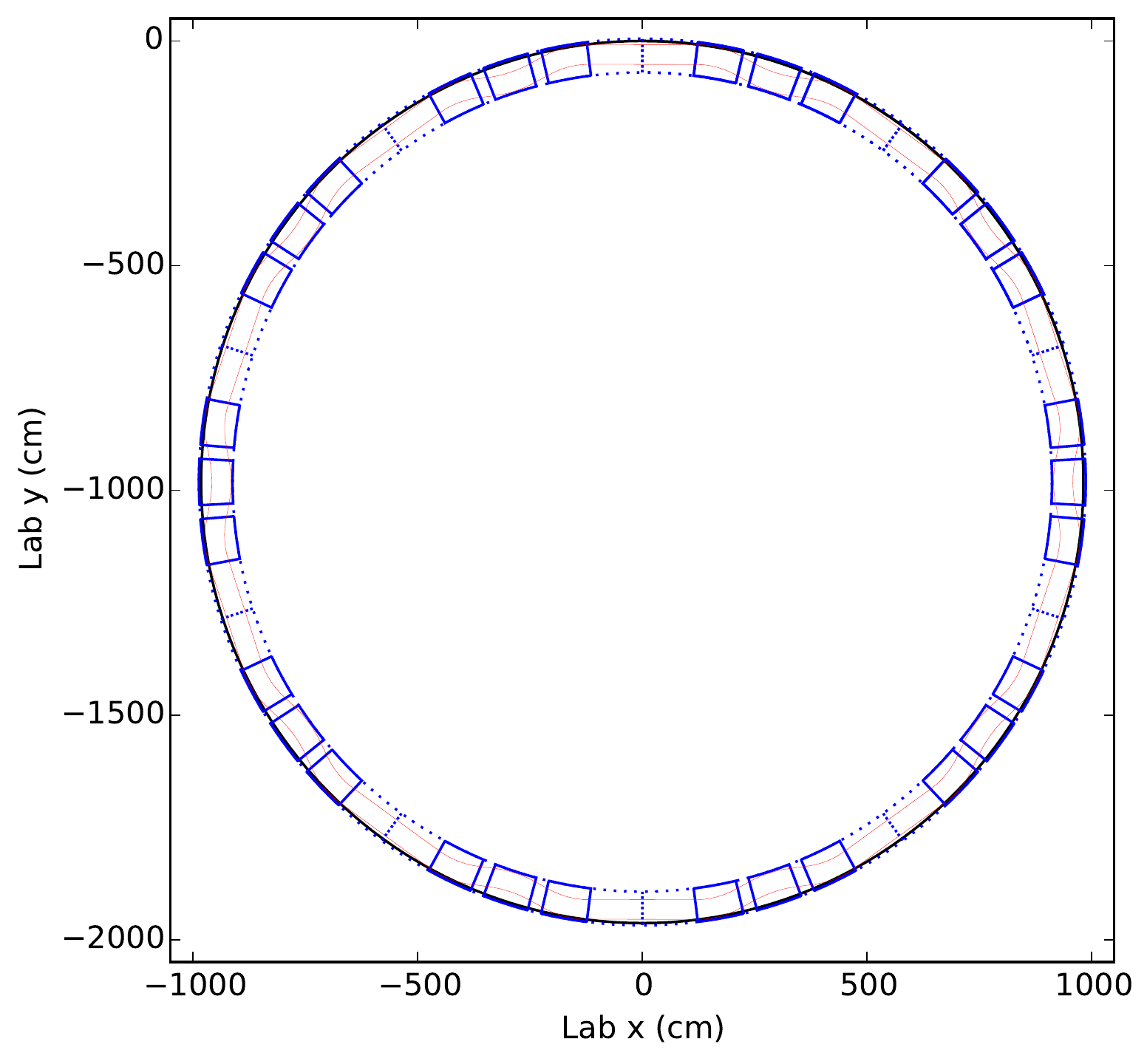}
  \caption
{NORMA round lattice, with minimum and maximum closed orbits shown in red.}
\label{fig:norma_ring}
\end{figure}

A NORMA triplet cell is shown in Fig.~\ref{fig:norma_cell}. The full ring composed of 10 cells is shown in Fig.~\ref{fig:norma_ring}. Between each triplet is 2.4~m of magnet-free drift space.
The parameters for the round NORMA lattice are given in table~\ref{tab:NORMA_ringparams}. In this article we refer to this lattice as the nominal NORMA design and use it as the baseline for further study.

\begin{table}[!ht]
\caption{The main parameters of the nominal NORMA lattice.}
    \centering
    \begin{ruledtabular}
    \begin{tabular}{l r}
        \bf{Parameter} &    \\
        \hline
        Injection energy & 50~MeV \\
        Maximum energy & 350~MeV \\
         Average radius & 9.61~m  \\
         Circumference & 60.4~m  \\
         Average orbit excursion & 0.43~m \\
         Ring tune ($Q_x$, $Q_y$) & 7.72, 2.74 \\
         Field index & 27.47\\
         Number of cells & 10  \\
         Max. field in F magnet & 1.57~T\\
         Max. field in D magnet & -1.19~T\\
         Dynamic aperture (normalized) &  68.0 mm\,mrad\\
         Magnet-free drift $L_{LD}$ &  2.4~m \\
    \end{tabular}
    \end{ruledtabular}
\label{tab:NORMA_ringparams}
\end{table}

Note that --- during the early stages of this work --- with idealized and 2D magnets we considered an energy range down to 30~MeV. Some of the dynamic aperture simulations in sections \ref{sec:2d_magnets} and \ref{sec:fringe_fields} were carried out at 30~MeV, however this is always a tougher requirement than at higher energy due to adiabatic damping.

\subsection{Tracking simulations}
\label{sec:magnet_tracking_models}

Zgoubi, which was used for tracking simulations, is a charged-particle tracking code widely used for designing and studying FFAGs. It uses a stepwise ray-tracing method; the particle is propagated in small steps and at each of these the magnetic field and its derivatives are evaluated. This gives accurate results for particles over a wide range of momenta and trajectories when moving through large-aperture magnets. Zgoubi features a range of magnet descriptions that together are capable of simulating the complex magnets which are typical of FFAGs. We use the PyZgoubi~\cite{pyzgoubi, tygier2015pyzgoubi} framework around Zgoubi to expand its capability, allowing advanced scripting and optimization.

Zgoubi features both analytic and field-map-based magnet descriptions. In each, the magnet description is used to calculate the field and its derivatives at each integration step along the charged-particle's trajectory. In our studies we have used several magnet models: \texttt{FFAG}, an analytic model of a sector scaling FFAG magnet; \texttt{DIPOLES}, a sector dipole with optional higher multipole fields; \texttt{POLARMESH}, a 2D polar mid-plane field map with an out-of-plane expansion.

\subsubsection{Analytic Scaling FFAG}

Zgoubi offers an idealized scaling FFAG sector magnet, which we use as the reference model for the nominal error-free lattice. The field is composed of the product of the radial scaling law
\begin{equation} \label{eq:scaling_law}
B_N = B_0 (r/r_0)^k,
\end{equation}
where $r_0$ is the reference radius and $k$ is the field index, and a longitudinal fringe function
\begin{equation} \label{eq:enge_fringe}
F = \frac{1}{1+\exp\left( C_0 + C_1\left(\frac{s}{\lambda}\right) + C_2\left(\frac{s}{\lambda}\right)^2 + \ldots \right)}
\end{equation}
where $s$ is the distance from the effective field boundary, $\lambda$ is the fringe field extent and $C_i$ are the well-known Enge coefficients~\cite{enge1964effects}. In the nominal design we use $C_1=$~2.24 and $\lambda=$~4~cm with other coefficients set to zero. Zgoubi allows any field overlap between magnets to be modelled by assuming linear superposition.

We use the \texttt{FFAG} element for initial optimization of the lattice; however it is limited in its flexibility for error studies as only the position and strength can be adjusted.

\subsubsection{Multipole expansion}

Zgoubi's \texttt{DIPOLES} element can be used to model a sector dipole, but also allows additional multipole components in the field. This can be used to describe combined-function magnets, but also to approximate more complex fields such as a scaling FFAG. It allows the radial field profile to be expressed as
\begin{equation} \label{eq:dipoles_field}
B_N = B_0 + B_1 (r-r_0) + B_2 (r-r_0)^2 + \ldots
\end{equation}
where $B_i$ are the multipole components. Again this is multiplied by the longitudinal fringe field given in Eq.~\ref{eq:enge_fringe} to give the mid-plane field.

\begin{figure}[htbp]
  \centering
  \includegraphics[width=0.9\columnwidth]{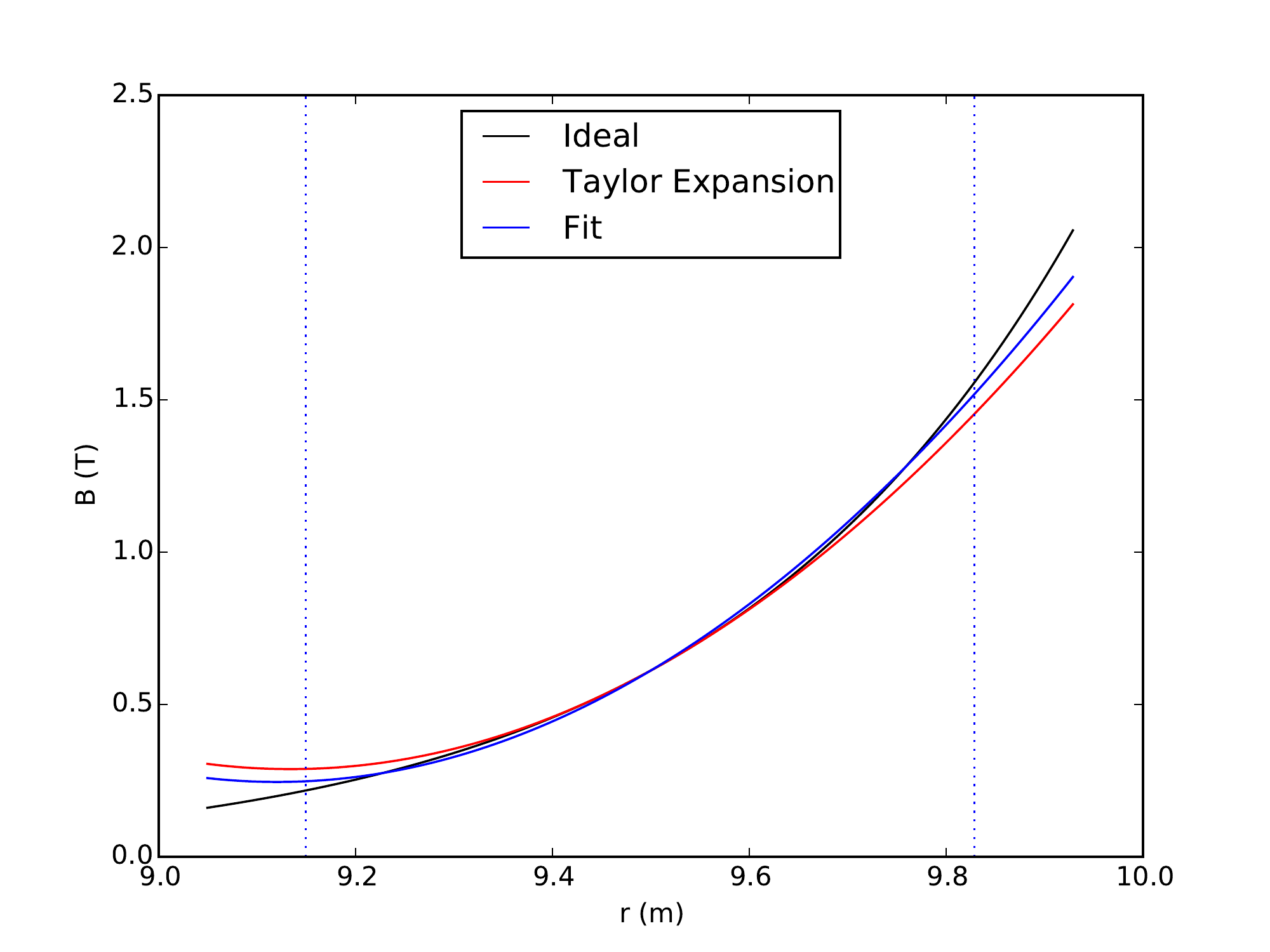}
  \caption
{Multipole Taylor expansion and fit to an example analytic scaling field; both with dipole, quadrupole and sextupole terms.}
\label{fig:multipole_expansion}
\end{figure}

There are two methods to find the multipole coefficients to use with the DIPOLES element. The ideal FFAG scaling field can be Taylor expanded about a given radius, usually halfway between the minimum and maximum orbits. Alternatively a multipole expansion can be fitted over the required good field region.
A Taylor expansion of $B(r)$ around $R_0$ gives
\begin{align} \label{eq:scaling_law_exp}
B_N = B_0 \left( 1 +
             \frac{k}{R_0} (r{-}R_0) + 
             \frac{k(k{-}1)}{2! R_0^2} (r{-}R_0)^2 +{} \right.\nonumber\\
    \left.   \frac{k(k{-}1)(k{-}2)}{3! R_0^3} (r{-}R_0)^3 + \ldots
\right),
\end{align}
which can be equated with Eq.~\ref{eq:dipoles_field} to obtain the coefficients.
Whilst the Taylor expansion gives the correct field and derivatives about the expansion point, using a fit gives a lower maximum deviation in field for any given expansion order, as illustrated in Fig.~\ref{fig:multipole_expansion}.
Use of \texttt{DIPOLES} with a fit up to \nth{9} order gives a very good agreement to the dynamics of the \texttt{FFAG} element: the mean closed orbit deviation is below $10^{-9}$~cm and the mean fractional deviation of the tune is below $10^{-10}$ horizontally and below $10^{-5}$ vertically.

\subsubsection{Midplane field maps}

Zgoubi's \texttt{POLARMESH} element allows a magnet to be defined in terms of a 2D polar field map in the mid-plane, with field maps generated in external magnet simulation codes such as OPERA~\cite{Opera}. This allows simulation of the deviations from an ideal field that are likely to occur in a real magnet.
It is also possible to generate field maps representing the ideal field by evaluating the analytic equations for a scaling FFAG magnet at the grid points. This can be used to distinguish simulation effects due to an interpolation step from actual effects due to the imperfections of a realistic magnet. It also allows investigation into the resolution requirements for the field map.
We find that for the nominal NORMA lattice a 1000$\times$1000 grid of mesh points per half cell (from the start to the center of the triplet) is sufficient to give a mean closed orbit deviation of $~10^{-6}$~cm and mean fractional tune deviations of $~10^{-5}$ horizontally and $~10^{-4}$ vertically. Increasing the mesh density to 2000$\times$2000 does not significantly improve agreement with the analytic models.

\subsection{Dynamic aperture}
\label{sec:dynamic_aperture}

Dynamic aperture (DA) is a measure of the stable area of phase-space for particles circulating in an accelerator. A particle within the stable area will survive a large number of turns though the accelerator lattice, while a particle outside this region will be lost after a small number of turns. In practice a finite number of turns must be simulated in order to predict the stability. For NORMA we consider a coordinate stable if a particle starting there will survive 1000 turns, as this is representative of the length of the acceleration cycle of around 1~ms. We use a strict definition for dynamic aperture where - for any given amplitude - a set of particles with a range of phase-space angles are tested and must all be stable, as described in~\cite{tygier2015pyzgoubi}.

A large dynamic aperture is required to transport the injected bunches through the accelerator with a low loss. A large dynamic aperture increases overall transmission efficiency (from injection to extraction) and therefore reduces radiation and activation. The injected bunches from the cyclotron will have a typical normalized emittance of less than 10~mm\,mrad~\cite{sabaiduc}. We therefore require that the normalized dynamic aperture is kept around 50~mm\,mrad or greater, a specification considered sufficient for this application~\cite{peach2013pamela}.

\begin{figure}[htbp]
  \centering
  \includegraphics[width=0.9\columnwidth]{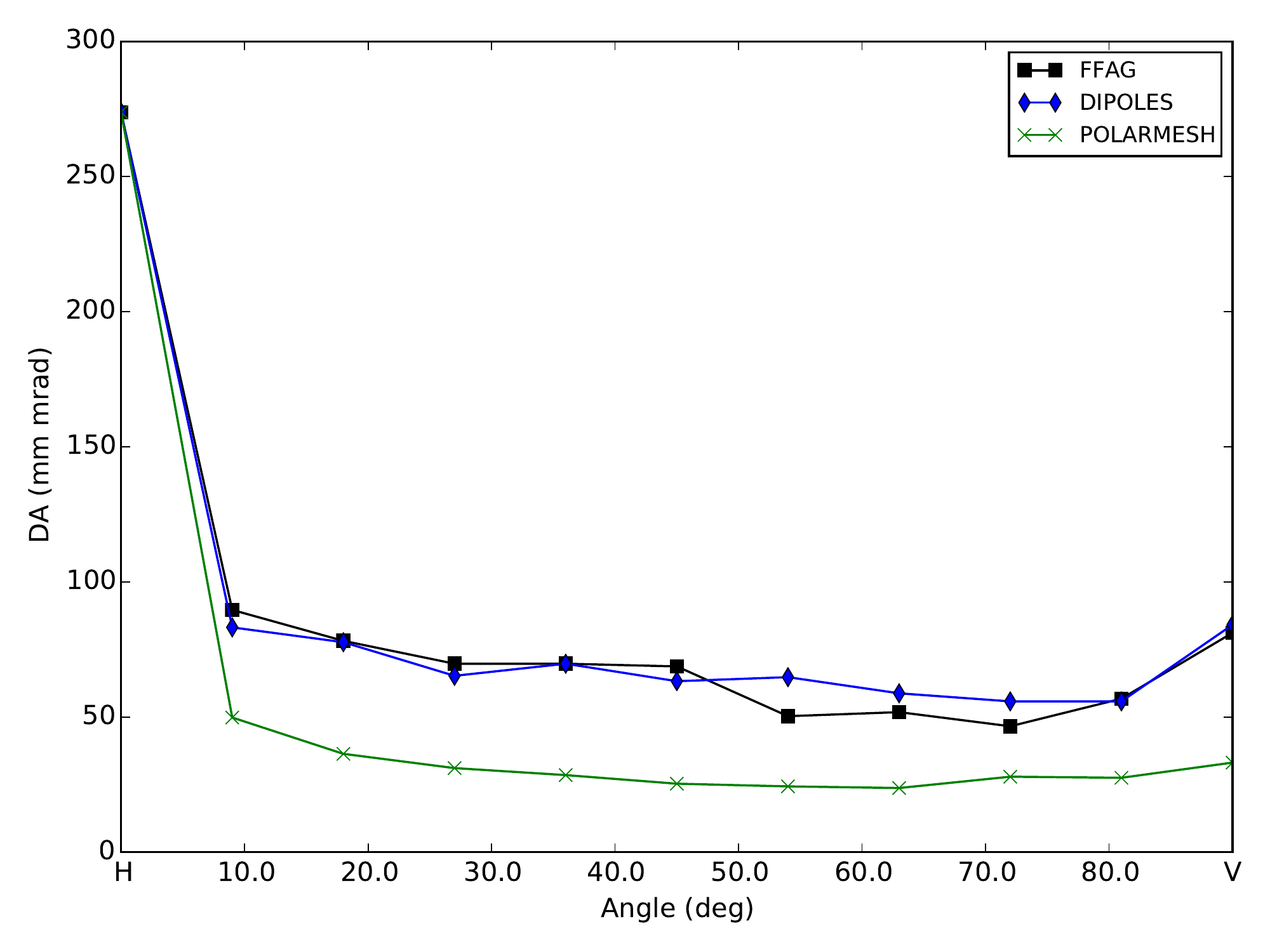}
  \caption
{Dynamic aperture as a function of real space angle for the nominal design modelled with FFAG, DIPOLES and POLARMESH elements.}
\label{fig:compare_da}
\end{figure}

It is useful to see the effect on the dynamic aperture of the nominal lattice design from using the differing tracking methods, before studying modifications to the lattice. This allows us to distinguish DA changes that are due to using a different field from those due to the choice of tracking method.
Figure~\ref{fig:compare_da} shows the dynamic aperture over a range of real space angles in $x$ and $y$, i.e. where 0\degree{} is the horizontal and 90\degree{} is the vertical dynamic aperture, using \texttt{FFAG}, \texttt{DIPOLES} and \texttt{POLARMESH} elements. As before, the \texttt{DIPOLES} shows good agreement to the \texttt{FFAG} element. However, with \texttt{POLARMESH} we see a significant reduction in DA. Increasing the number of mesh points used in the \texttt{POLARMESH} does not improve agreement with the analytic models. We believe that this is due to small errors from the interpolation of the field map building up when tracking for a large number of turns. Therefore, we consider only \texttt{FFAG} and \texttt{DIPOLES} elements to be suitable for DA calculations.

\subsection{Field Errors}
\label{sec:errors}

Deviations of a given field from an ideal field can be measured in a number of ways, and it is common to specify maximum deviations from the ideal field or gradient. In general the deviation of each multipole component from the ideal field can be measured. In a synchrotron a multipole expansion about the magnet center can be used, but for a magnet that accepts a wide range of orbits one needs to be careful about where the multipole components are measured. This is important because introducing a multipole error of a given order at one location will change the lower-order multipole values everywhere else; for example a sextupole error at a given orbit radius causes a quadrupole and dipole shift across the magnet. For a field map defined by field strengths on a regular grid, the multipole components can be found either by fitting a polynomial to the whole map or a subsection of it, or by repeated numerical differentiation (e.g. the quadrupole component is proportional to the \nth{1} derivative of the field with respect to the radius).

\newcommand{\bnom}{\ensuremath{B_{N}}}
\newcommand{\bact}{\ensuremath{B_{A}}}
\newcommand{\magb}{\ensuremath{B_{0}}}
\newcommand{\magr}{\ensuremath{r_{0}}}
\newcommand{\magk}{\ensuremath{k}}

\section{Magnet design and optimization}
\label{sec:magnet_design_simulations}

This section describes the finite element models of the lattice magnets, the strategy chosen for their optimization and the main steps of its implementation.

\subsection{\label{sec:2D}2D magnet models}
\begin{figure}[htbp]
\includegraphics[width=8.5 cm]{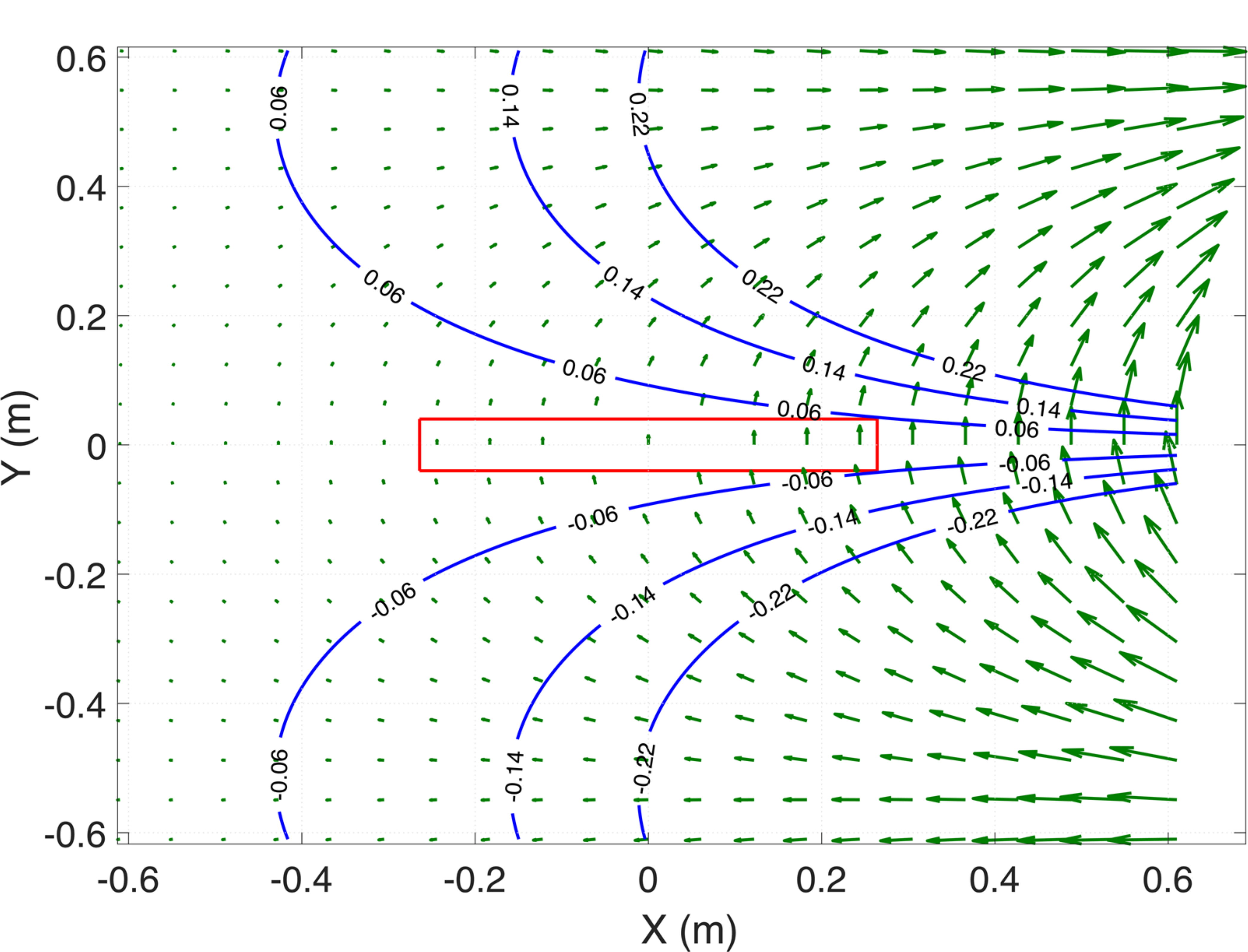}
\caption{Lines of constant scalar potential for the D-magnet (blue). The good-field region required is represented by the rectangle (red) and the origin of the coordinate system is the center of the good-field region. The vectors (green) are proportional to the local gradient of the scalar potential.}
\label{fig:fig1kiril}
\end{figure}

The required nominal radial field  profile $B_{N}(r)$ is given by Eq.~\ref{eq:scaling_law}.
The values of the parameters $r_{0}$, $B_{0}$, $k$ and the extent of the good-field region required are specified for each magnet. The first step of the magnet design and optimization process is to fit Eq.~\ref{eq:scaling_law} to a polynomial within the good field region. By following the standard procedure (see e.g.~\cite{Brandt09}) the coefficients of the fitting polynomial can be used to obtain the two-dimensional scalar potential $\Psi  = \Psi \left( {x,y} \right)$  such that ${\bf{B}} = \nabla \Psi $  in the air gap of the two magnets. Figure~\ref{fig:fig1kiril} shows the result. Note, that the 2D magnet modelling and optimization is performed in a coordinate system with an origin located at the center of the good-field region of each magnet and the $x$-axis points along the machine radius.

If the permeability of the magnet yoke material is infinitely high the lines of constant scalar potential in Fig.~\ref{fig:fig1kiril} coincide with the faces of the magnet poles that generate the required field. The good-field region (plus any reasonable contingencies) must fit within the air gap of the magnet and this determines uniquely the actual magnet pole shapes and the nominal value of the scalar potential that the magnet will be operated at. Indeed, as Fig.~\ref{fig:fig1kiril} shows, each pole configuration is uniquely determined by the absolute value of its scalar potential $\left| {{\Psi _0}} \right|$. Finally, the nominal magnet current ${I_N} = {{2\left| {{\Psi _0}} \right|} \mathord{\left/
 {\vphantom {{2\left| {{\Psi _0}} \right|} {{\mu _0}}}} \right.
 \kern-\nulldelimiterspace} {{\mu _0}}}$  (Ampere-turns) can be obtained from Amp\`{e}re's law as usual, where $\mu_{0}$ is the vacuum permeability.

However, in reality the permeability of ferromagnetic steel normally employed in accelerator magnet construction is finite. Therefore, strictly speaking the magnet pole shape obtained by means of the multipole expansion technique described in the preceding paragraph is an approximation only. In addition, the onset of magnetic steel saturation lowers the permeability even further and this leads to field errors that cannot be neglected, particularly in the high-field regions of the magnets. 
If the field error $\Delta B = {B_N} - {B_M}$, where ${B_M} = {B_y}(x,y = 0)$  is the actual field generated by the magnet, is too high then the pole shape $y = y(x)$  must be replaced with a new pole shape $\tilde y = y(x) + \Delta y(x)$ such that the magnitude of the field error ${B_N} - {\widetilde B_M}$  it generates is acceptable. The approximate relationship between $\Delta B$  and $\Delta y(x)$  can be obtained as follows. Since both ${B_N}$  and ${B_M}$  are functions of $x$ i.e. ${B_N} = {B_N}\left( x \right)$  and ${B_M} = {B_M}\left( x \right)$  one can substitute $x = x\left( y \right)$  in the two expressions by inverting the known expression for the pole profile $y = y\left( x \right)$  and obtain the new expressions ${B_N} = {B_N}\left( y \right)$  and  ${B_M} = {B_M}\left( y \right)$. Naturally, this is only possible under the assumption that the inverse pole profile function exists. The next step is to express ${B_N}$  in the form
\begin{equation}\label{eq2}
{B_N} \approx {B_M} + \frac{{d{B_M}}}{{dx}}\frac{{dx}}{{dy}}\Delta y + \frac{1}{2}\frac{d}{{dy}}\left( {\frac{{d{B_M}}}{{dx}}\frac{{dx}}{{dy}}} \right)\Delta {y^2} + ...
\end{equation}
where $\Delta y\left( x \right)$  is the correction to the pole profile. In Eq.~\ref{eq2} terms of the order of $\Delta {y^3}$  and higher have been neglected. An approximate solution to Eq.~\ref{eq2} is easy to obtain and the result is
 \begin{equation}\label{eq3}
\Delta y \approx {\alpha _1}\left( {{B_N} - {B_M}} \right)\left( {1 - 0.5{\alpha _2}\left( {{B_N} - {B_M}} \right) + ...} \right)   
 \end{equation}
where
\begin{equation*}
    {\alpha _1} = \frac{{dy}}{{dx}}{\left( {\frac{{d{B_M}}}{{dx}}} \right)^{ - 1}}
\end{equation*}
and
\begin{equation*}
    {\alpha _2} = {\left( {\frac{{d{B_M}}}{{dx}}} \right)^{ - 1}}\left[ {\frac{{{d^2}{B_M}}}{{d{x^2}}}{{\left( {\frac{{d{B_M}}}{{dx}}} \right)}^{ - 1}} - \frac{{{d^2}y}}{{d{x^2}}}{{\left( {\frac{{dy}}{{dx}}} \right)}^{ - 1}}} \right].
\end{equation*}
In Eq.~\ref{eq3} terms of the order of ${\left( {{B_N} - {B_M}} \right)^3}$  and higher have been neglected. Equivalently Eq.~\ref{eq3} can be derived by formally inverting the function ${B_M} = {B_M}\left( y \right)$  and Taylor-expanding the result $y = y\left( B \right)$. 

It is clear that $\frac{dB_{M}}{dx}$ must not be zero as it is present in the denominators of both the expressions for $\alpha_{1}$ and $\alpha_{2}$.  However, in deriving Eq.~\ref{eq3} it has been assumed that the term of the order of ${\left( {{B_N} - {B_M}} \right)^2}$  is a small perturbation compared to the term of the order of $\left( {{B_N} - {B_M}} \right)$, (or equivalently, the term proportional to $\alpha_2$ in Eq.~\ref{eq3} is much smaller than one) and so the series can be truncated without any significant loss of accuracy. This assumption may not be valid if the value of $\left| {\frac{{d{B_M}}}{{dx}}} \right|$   is sufficiently small. This means that in the vicinity of local extrema of $B_{M}$ Eq.~\ref{eq3} becomes inaccurate. In addition $\frac{{dy}}{{dx}}$  must be non-zero in order to ensure the existence of the inverse pole profile function $x=x(y)$. If however, $\left| {\frac{{dy}}{{dx}}} \right| \to \infty$ (i.e. $\frac{dy}{dx}$ has a pole (see Fig.~\ref{fig:fig1kiril}) then in the vicinity of that pole $\left| {\frac{{{d^2}y}}{{d{x^2}}}{{\left( {\frac{{dy}}{{dx}}} \right)}^{ - 1}}} \right| \to \infty $ and the term proportional to $\alpha_2$ in  Eq.~\ref{eq3} may no longer be small compared to one. Hence, the applicability of Eq.~\ref{eq3} relies upon the following two conditions: (i) $\left|\frac{{dy}}{{dx}}\right|$ is non-zero and not too large and (ii) $\left| {\frac{{d{B_M}}}{{dx}}} \right|$ is not too small.

\begin{figure}[htbp]
\includegraphics[width=8.5 cm]{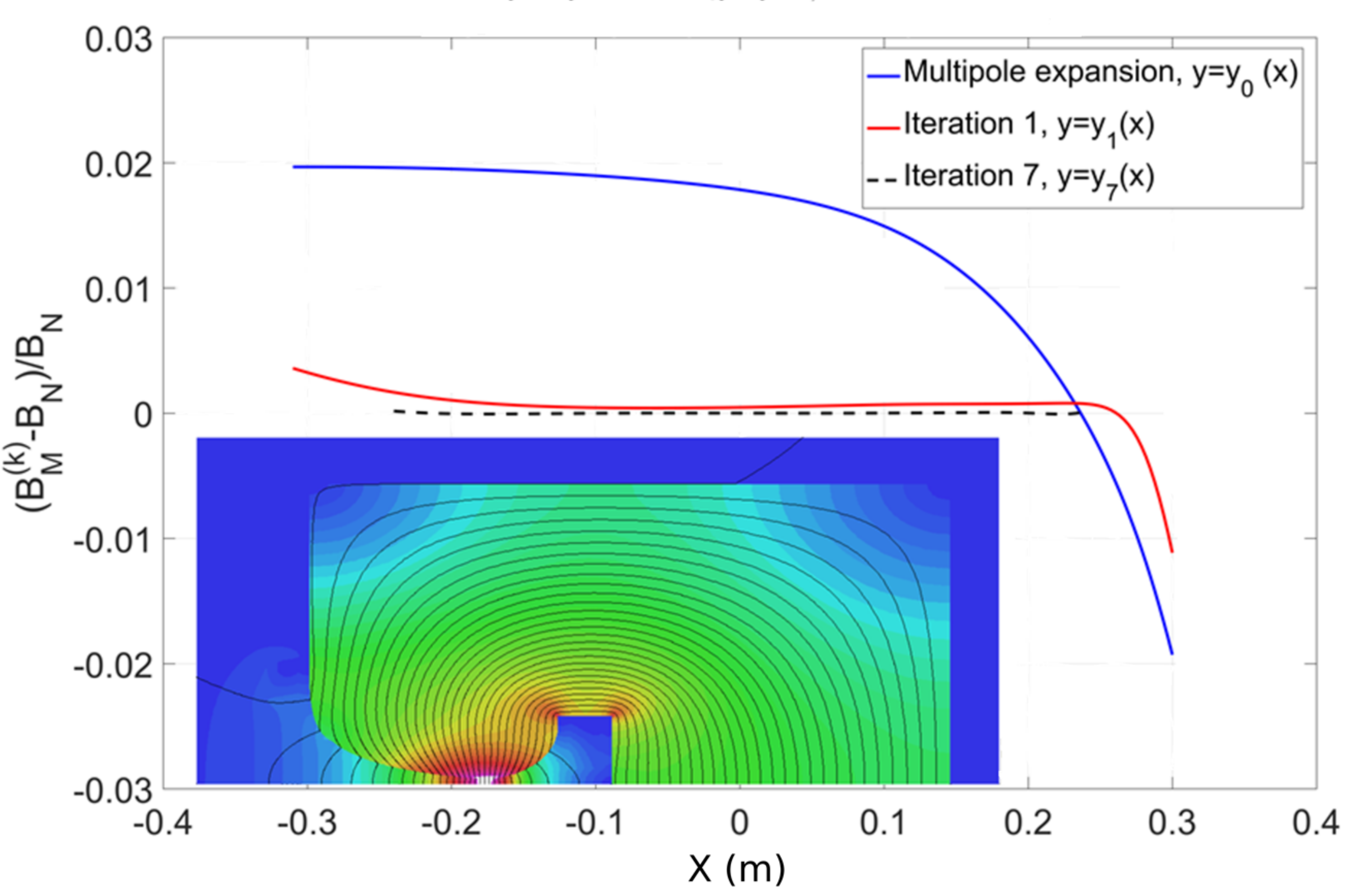}
\caption{A test of the pole optimization procedure performed on the 2D F-magnet model. Distribution of the relative field error obtained from three different iterations. The inset shows the 2D flux density distribution as obtained with OPERA 2D illustrating the low-field and high-field regions of the magnets. Red color corresponds to high flux density.}
\label{fig:fig2kiril}
\end{figure}

\begin{figure}[htbp]
    \centering
    \begin{subfigure}[b]{0.49\columnwidth}
        \includegraphics[width=\columnwidth]{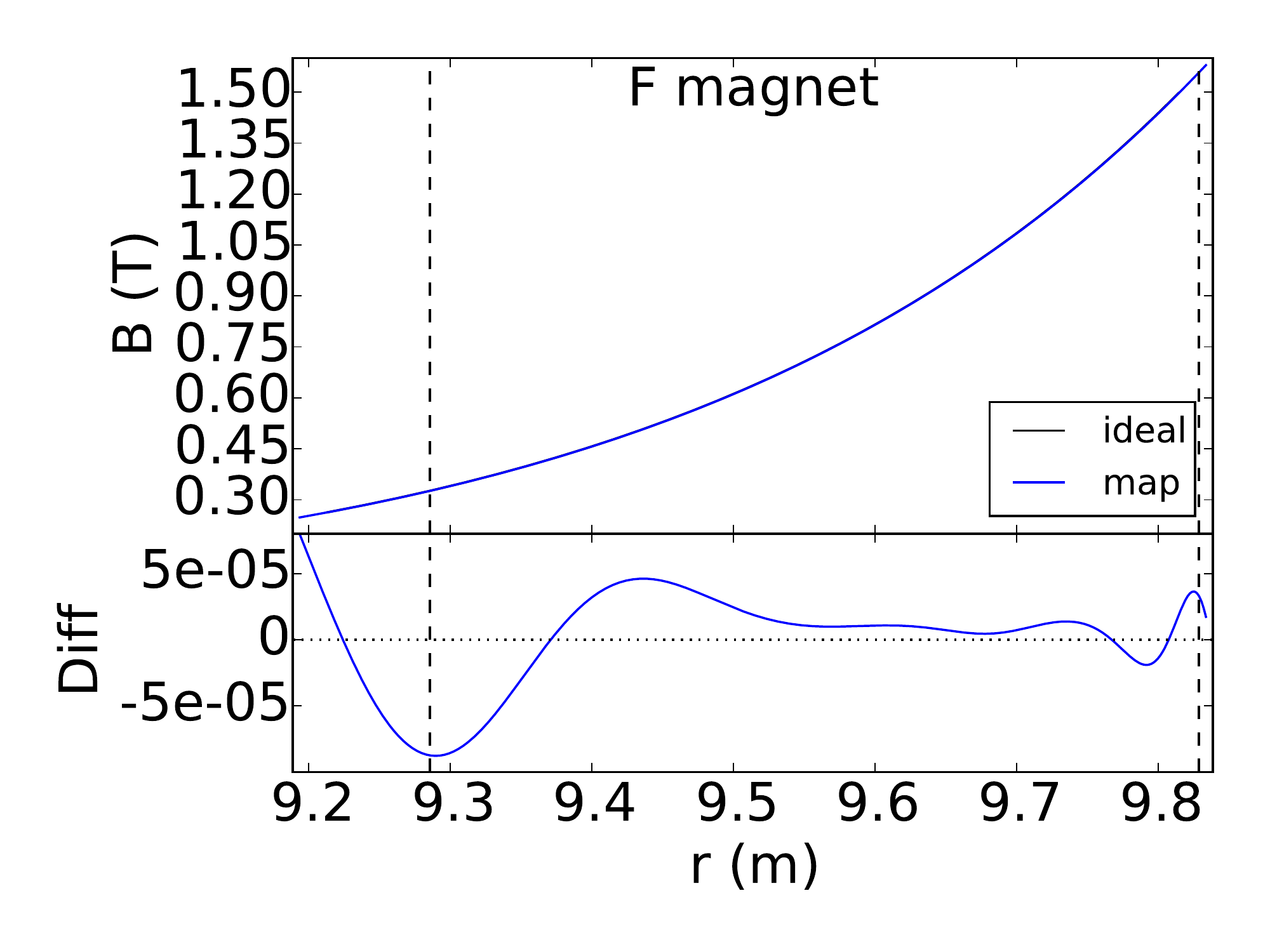}
        \caption{F dipole component}
        \label{fig:2d_profile_map_ideal_f_dipole}
    \end{subfigure}
    \begin{subfigure}[b]{0.49\columnwidth}
        \includegraphics[width=\columnwidth]{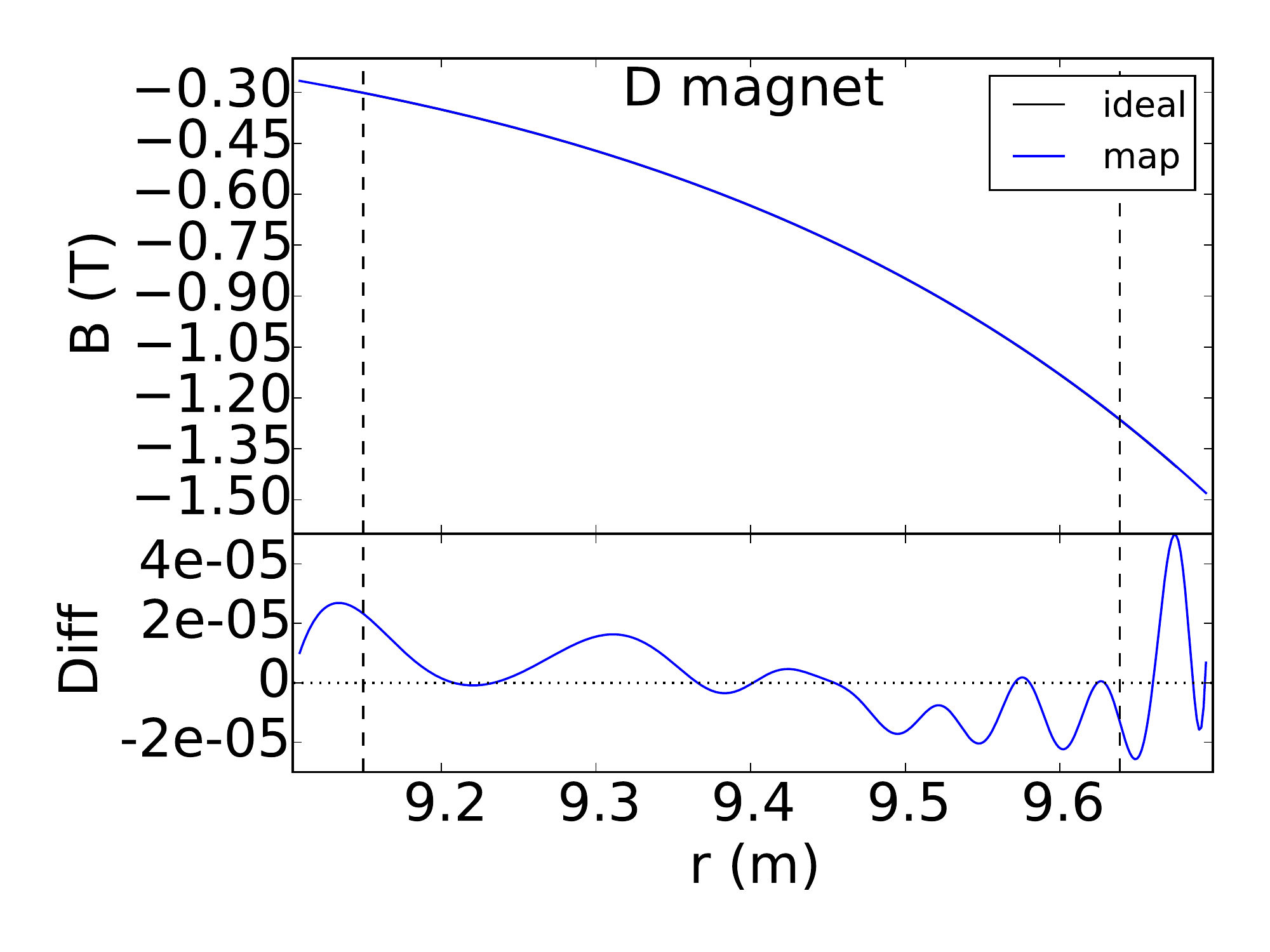}
        \caption{D dipole component}
        \label{fig:2d_profile_map_ideal_d_dipole}
    \end{subfigure}

    \begin{subfigure}[b]{0.49\columnwidth}
        \includegraphics[width=\columnwidth]{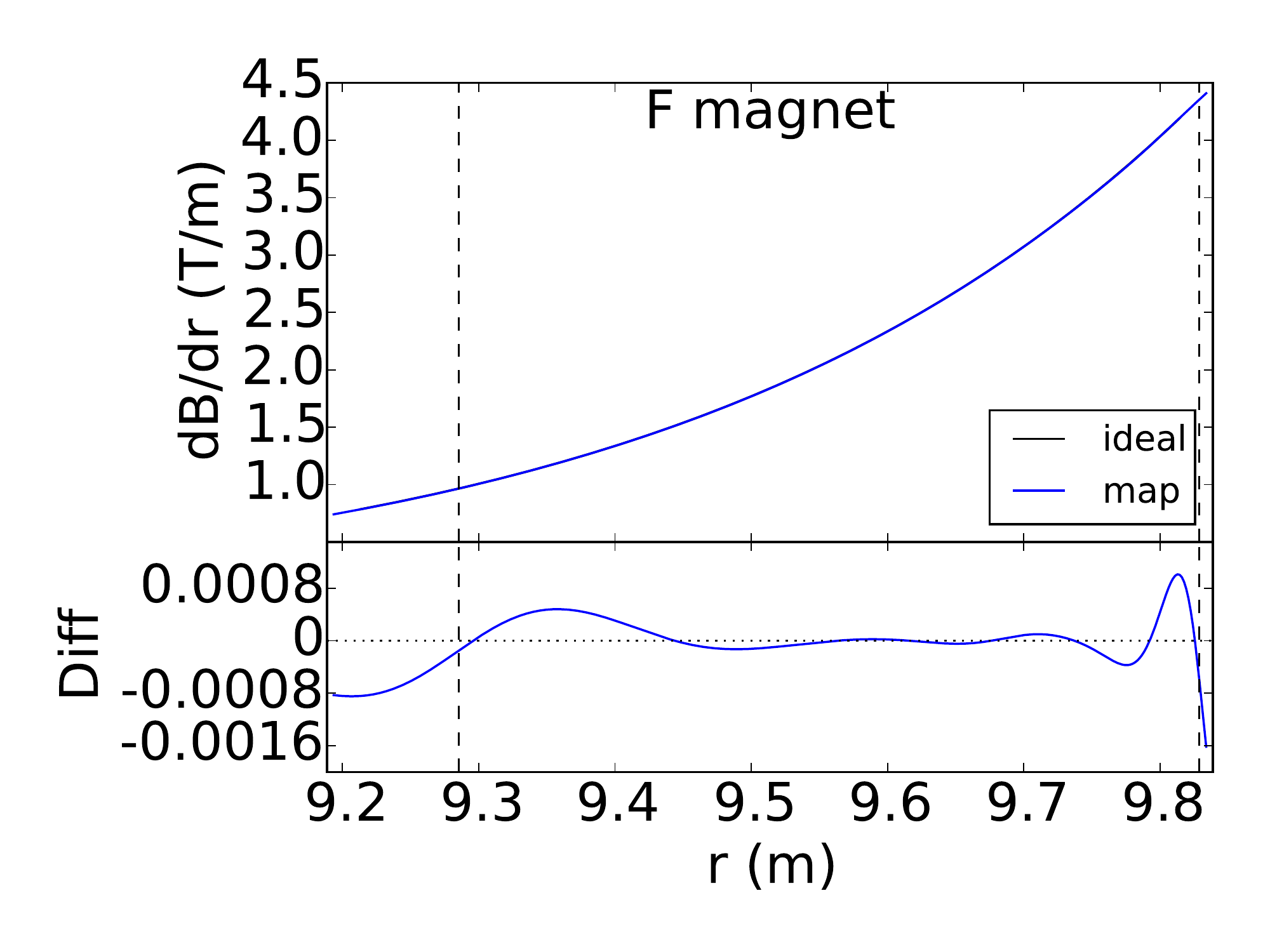}
        \caption{F quadrupole component}
        \label{fig:2d_profile_map_ideal_f_quadrupole}
    \end{subfigure}
    \begin{subfigure}[b]{0.49\columnwidth}
        \includegraphics[width=\columnwidth]{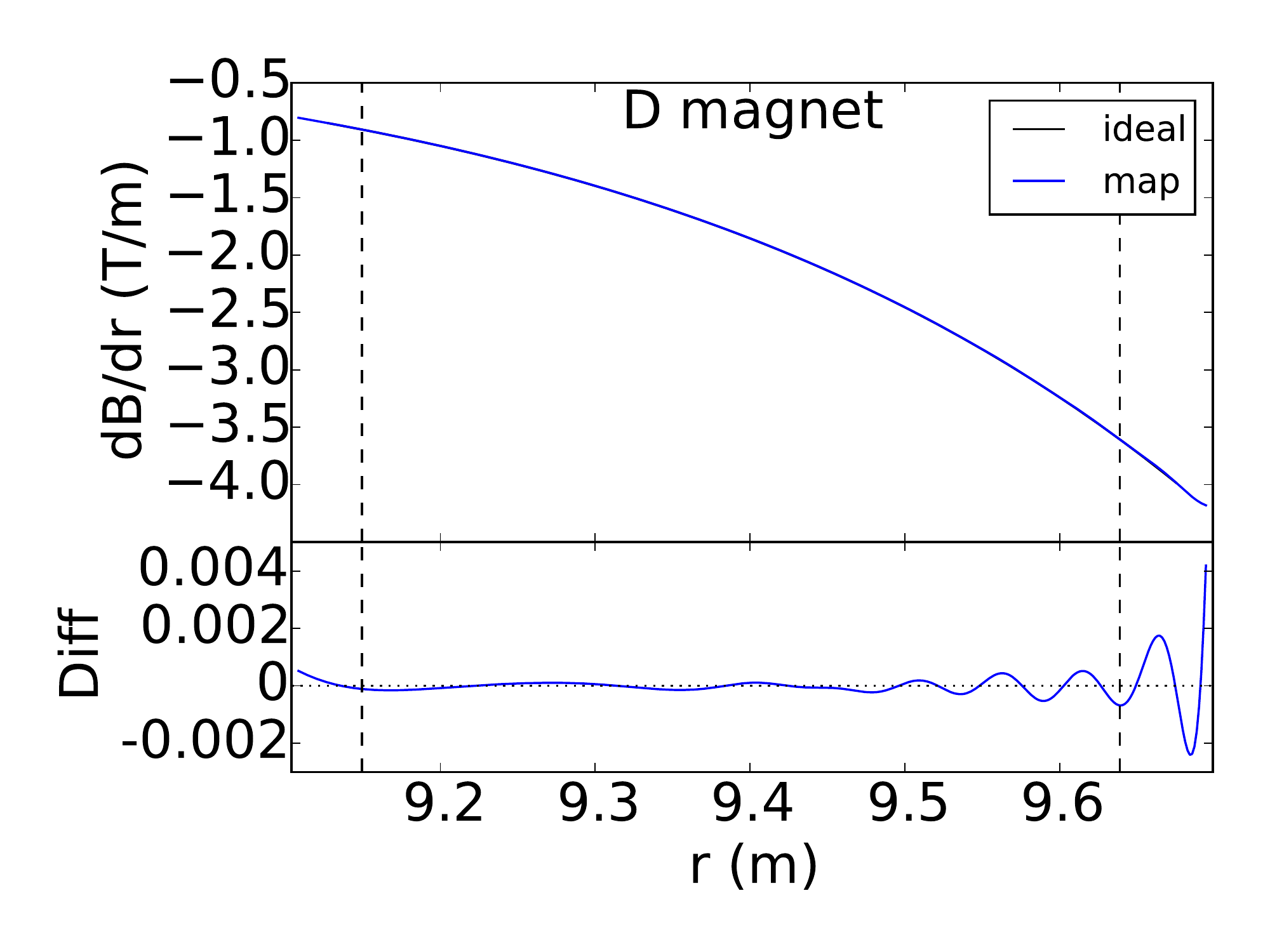}
        \caption{D quadrupole component}
        \label{fig:2d_profile_map_ideal_d_quadrupole}
    \end{subfigure}

    \begin{subfigure}[b]{0.49\columnwidth}
        \includegraphics[width=\columnwidth]{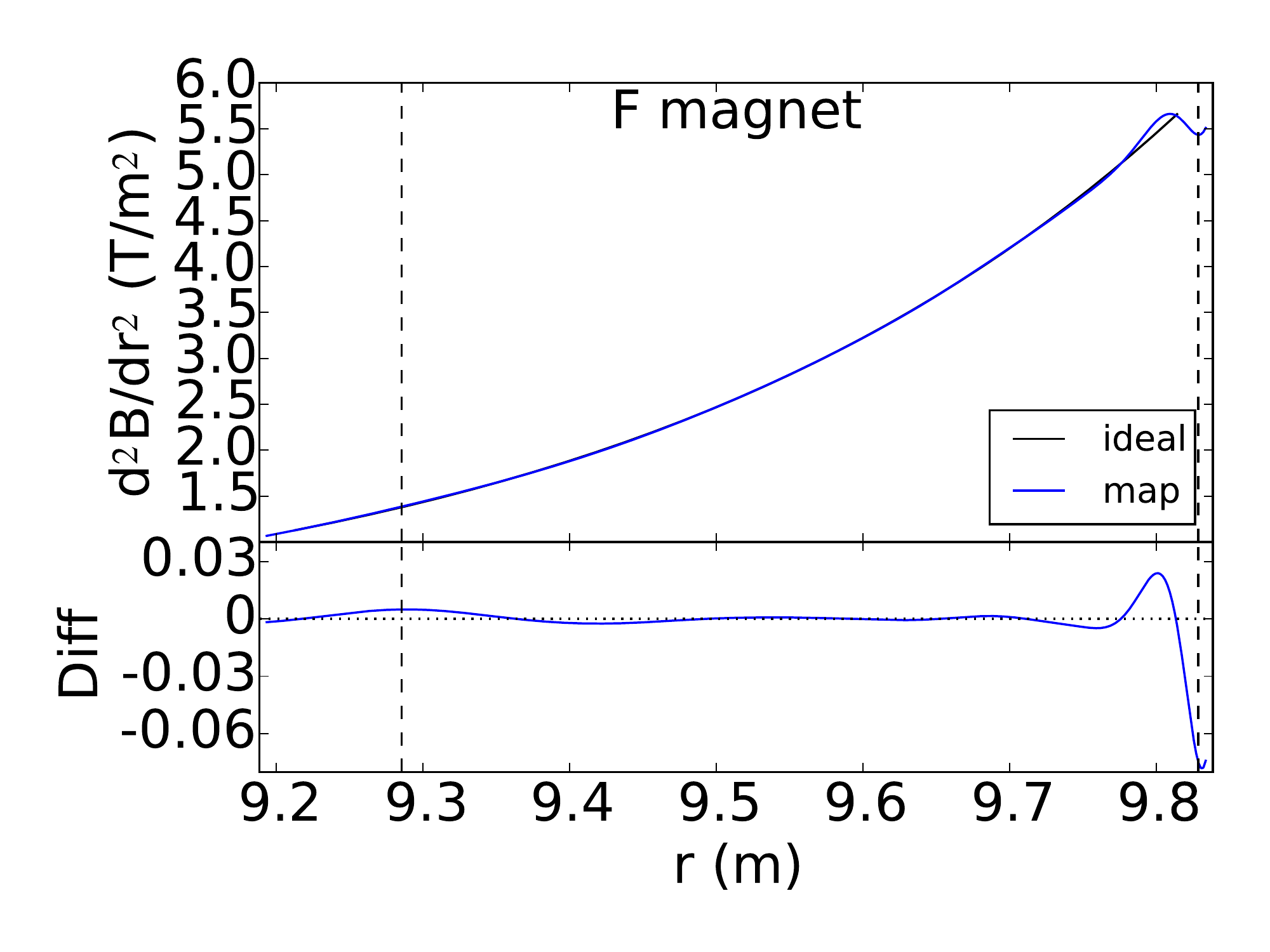}
        \caption{F sextupole component}
        \label{fig:2d_profile_map_ideal_f_sextupole}
    \end{subfigure}
    \begin{subfigure}[b]{0.49\columnwidth}
        \includegraphics[width=\columnwidth]{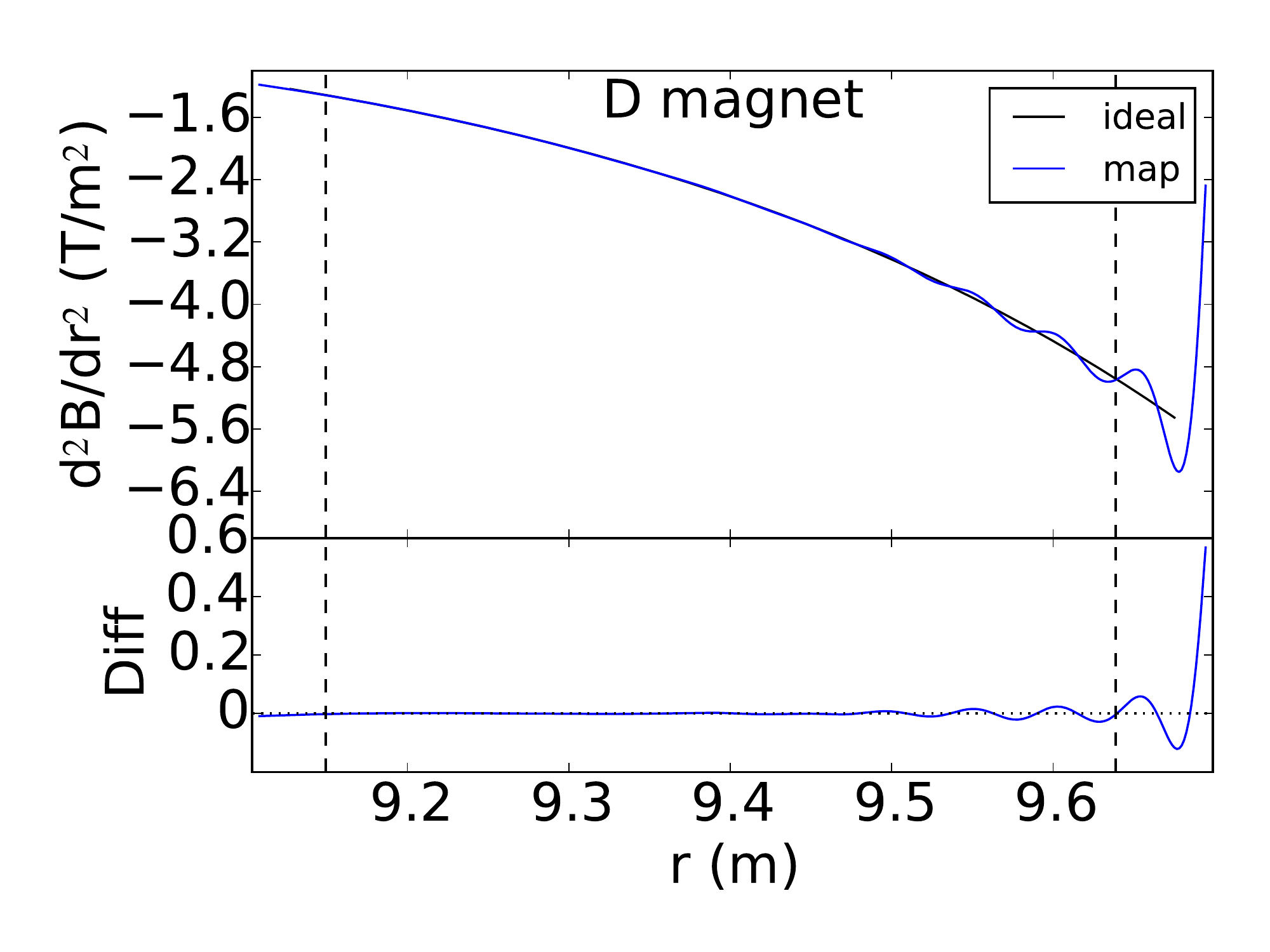}
        \caption{D sextupole component}
        \label{fig:2d_profile_map_ideal_d_sextupole}
    \end{subfigure}

    \caption{Multipole components of the 2D field profiles compared to the nominal field. $r$ is the radial distance from the machine center.}\label{fig:2d_profile_map_ideal}
\end{figure}

A magnet pole optimization procedure based on Eq.~\ref{eq3} must be applied iteratively as Eq.~\ref{eq3} itself is an approximation. The zeroth-order approximation to the pole profile ${y_0}\left( x \right)$ is obtained by means of the multipole expansion technique and the field distribution  $B_M^{(0)}\left( x \right)$ it generates is obtained from finite-element (OPERA 2D~\cite{Opera}) simulations. $B_M^{(0)}\left( x \right)$  and ${y_0}\left( x \right)$  are then substituted into the right-hand side of Eq.~\ref{eq3} to obtain the corrected pole profile  ${y_1} = {y_0} + \Delta {y_0}$ which is in turn used to obtain the field distribution $B_M^{(1)}\left( x \right)$  and so on. The implementation of this scheme on a computer uses MATLAB~\cite{Matlab} to calculate the pole profile correction according to Eq.~\ref{eq3} from field distribution data generated by OPERA 2D and the result is then fed to OPERA 2D, which in turn calculates the updated field distribution. This process is repeated until the desired maximum field error is reached. Figure~\ref{fig:fig2kiril} shows the result from a test run.
The initial (zeroth-order) pole profile and the current strength were obtained with the multipole expansion technique. In order to test the optimization algorithm the coil current was intentionally increased so that the field in the main part of the magnet (which operates far from saturation) is 2\% higher than the nominal field. Within 7 iterations the peak relative field error was reduced by a factor of over 200: from 2\% to less than $10^{-4}$. 
Figure~\ref{fig:2d_profile_map_ideal} shows the good match for the dipole, quadrupole and sextupole components between the 2D field profiles and the ideal nominal magnet field.

The first four iterations were performed in the region $-0.3 \hspace{.25em}\textrm{m} \hspace{.25em}<\hspace{.25em} x \hspace{.25em}<\hspace{.25em} 0.3 \hspace{.25em}\textrm{m}$ and the remaining three in the region $-0.25\hspace{.25em}\textrm{m}\hspace{.25em}<\hspace{.25em} x \hspace{.25em}<\hspace{.25em}0.25 \hspace{.25em}\textrm{m}$ while the specified good field region for this test was $-0.23 \hspace{.25em}\textrm{m}\hspace{.25em} < \hspace{.25em}x\hspace{.25em} <\hspace{.25em} 0.23 \hspace{.25em}\textrm{m}$. It was noted that the convergence in the central region of the magnet was faster than in the low-field and high-field regions. Indeed, as Fig.~\ref{fig:fig2kiril} shows, the extent to which the error is reduced by the first iteration is greater in the central part of the magnet than towards the magnet ends. A possible explanation for this is that in the low-field region $\frac{{dy}}{{dx}}$  gradually increases towards the pole roll-off area whilst in the high-field region $\frac{{d{B_M}}}{{dx}}$  approaches zero near the peak of the magnetic field. As pointed out earlier these are precisely the conditions when Eq.~\ref{eq3} is no longer accurate.
 
\begin{figure}[htbp]
\includegraphics[width=8.5 cm]{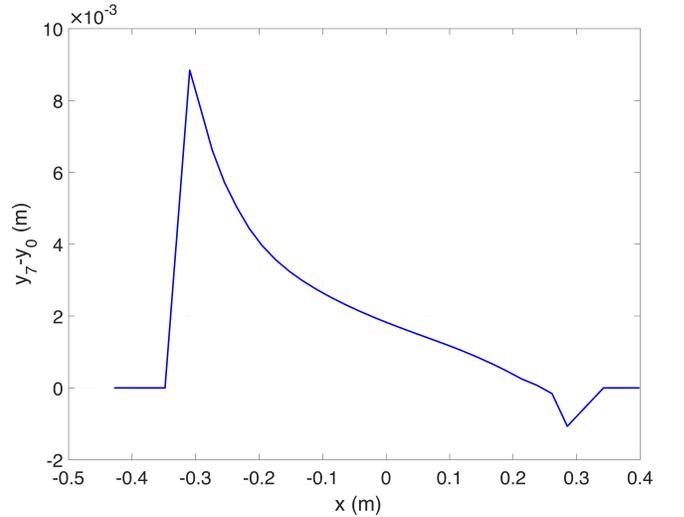}
\caption{Correction to the pole profile of the F-magnet obtained after 7 iterations.}
\label{fig:fig3kiril}
\end{figure}

Figure~\ref{fig:fig3kiril} shows the correction to the pole profile  obtained after 7 iterations. As can be seen, $\Delta y\left( x \right) \ge 0$  in the region  $-0.23\hspace{.25em} \textrm{m}\hspace{.25em}<\hspace{.25em} x\hspace{.25em} < \hspace{.25em}0.23\hspace{.25em} \textrm{m}$, which corresponds to increasing the magnet gap and decreasing the field strength in full accordance with Fig.~\ref{fig:fig2kiril}. The notch formed near $x=0.3 \hspace{.25em}\textrm{m}$ is most a likely a result of Eq.~\ref{eq3} becoming less accurate close to the region where  $\frac{{d{B_M}}}{{dx}} \to 0$.

\subsection{\label{sec:3D}3D magnet models}
\begin{figure}[htbp]
\includegraphics[width=8.5 cm]{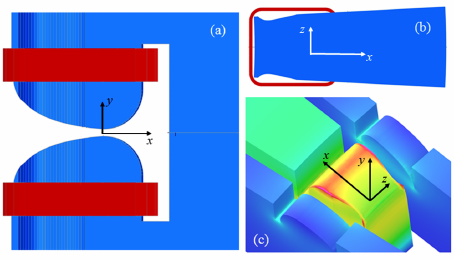}
\caption{3D model of the F magnet (a) Side view without the clamping plates (b) View from the top showing the pole edge profile without the clamping plates (c) Half of the pole area with the clamping plates. Their role is to limit the extent of the fringe fields and to eliminate cross-talk between adjacent magnets. The magnetizing coil is not shown.}
\label{fig:fig4kiril}
\end{figure}

The 3D magnet models were created by extruding the 2D magnet shapes, Fig.~\ref{fig:fig4kiril} shows the OPERA 3D model of the focusing (F) magnet. Two clamping plates were added on each side of the magnets to restrict the extent of their fringe fields and eliminate possible cross-talk between adjacent magnets. In order to save time the overall height of each magnet, the size of its back leg, the thickness of the clamping plates and the geometry of its roll-off in the high-field region were not optimized. The optimization in 3D was performed on the integrated field strength only and no attempt was made to adjust the main field (generated in the air gap of the magnet) and the fringe field (generated outside the air gap) separately. 

The integrated field strength $I(r)$ is defined as
\begin{equation}\label{eq4}
I\left( r \right) = \int\limits_{ - {\Gamma _0}}^{{\Gamma _0}} {{B_y}\left( {r,\varphi ,y = 0} \right)d\varphi } 
\end{equation} 					
where the origin of the cylindrical coordinate system is located in the machine center, and $r$, $\varphi$ and $y$ are the radial, azimuthal and axial coordinates, respectively, and ${B_y}\left( {r,\varphi ,y} \right)$  is the vertical (axial) magnetic field component. The integration limit $\Gamma_{0}>0$ was chosen such that ${B_y}\left( {r,\left| \varphi  \right| > {\Gamma _0},y = 0} \right) \to 0$. 

\begin{figure}[htbp]
\includegraphics[width=8.5 cm]{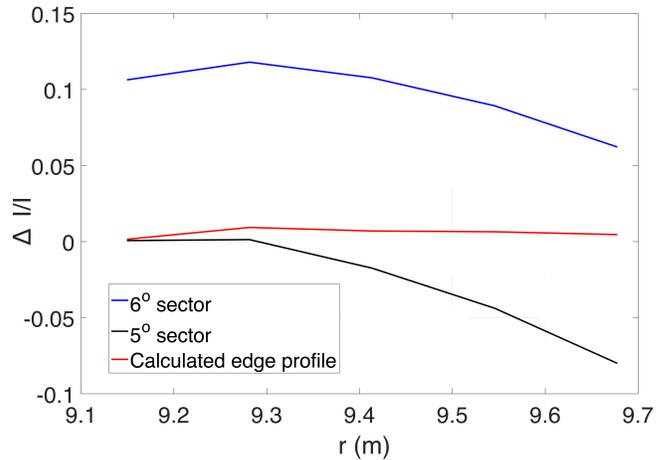}
\caption{Relative integrated field error for the F-magnet for a $6^\circ$-sector model, $5^\circ$-sector model and a model with an optimized edge. Here $r$ is the radial coordinate of the auxiliary cylindrical coordinate system $(r, \varphi, y)$ introduced.}
\label{fig:fig5kiril}
\end{figure}

\begin{figure}[htbp]
\includegraphics[width=8.5 cm]{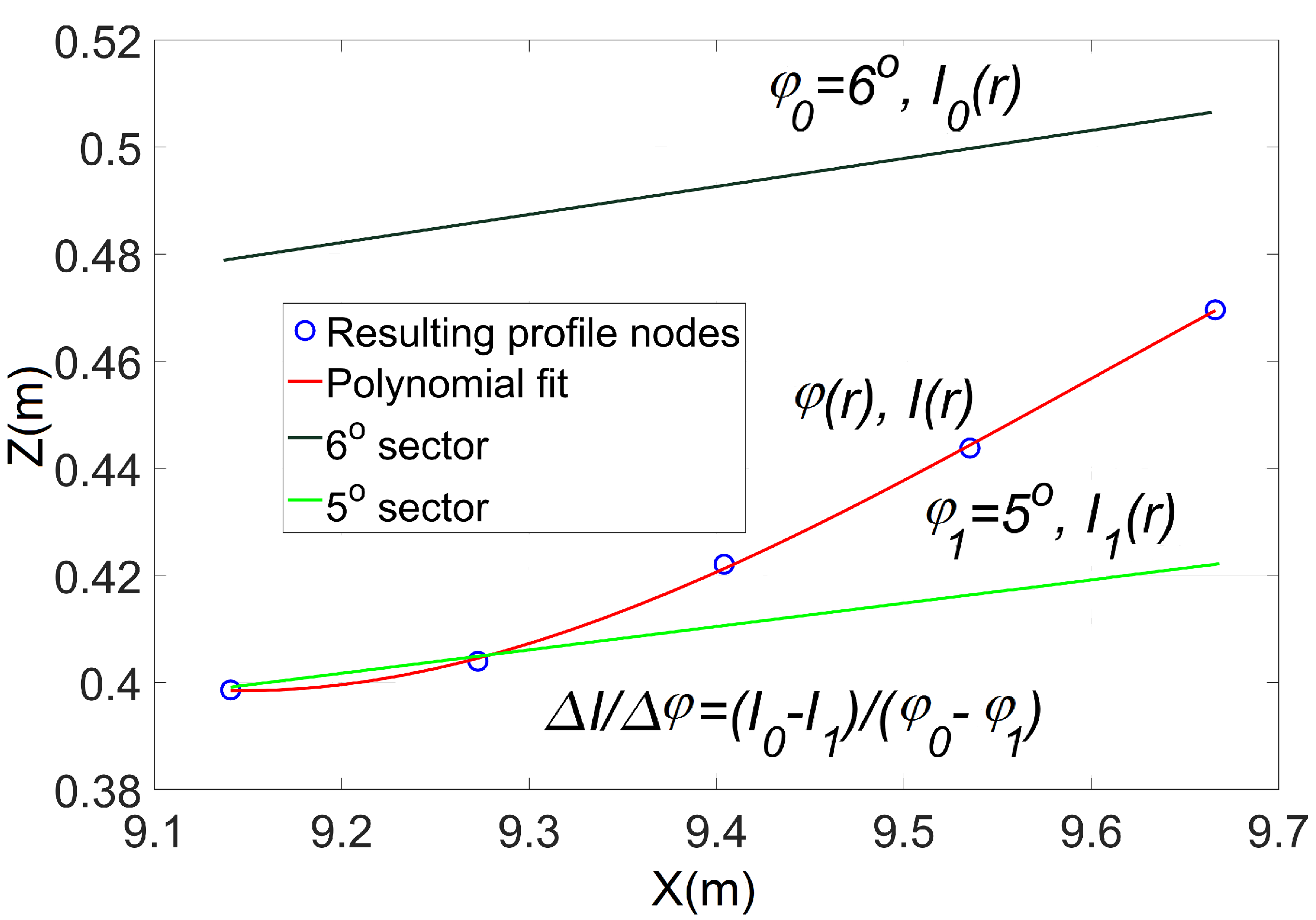}
\caption{The pole edge profile $\varphi=\varphi(r)$ of the F-magnet obtained from Eq.~\ref{eq5} is plotted in a Cartesian coordinate system $X=r\hspace{.25em} \textrm{cos}\varphi(r)$ and $Z=r\hspace{.25em} \textrm{sin}\varphi(r)$} and compared to a straight $6^\circ$-edge and a $5^\circ$-edge.
\label{fig:fig6kiril}
\end{figure}

Both magnets were specified as six-degree sector magnets. It was found that the integrated field strength distribution generated by the 3D six-degree sector magnet models with straight edges deviates considerably from the ideal, ``hard-edge'', six-degree sector magnetic field distribution given by Eq.~\ref{eq:scaling_law} (see Fig.~\ref{fig:fig5kiril}). This means that pole edges of the magnets need to be adjusted accordingly.
To achieve this the new pole edge shape $\varphi=\varphi(r)$ in the polar coordinate system employed is calculated from Newton-Raphson's formula
\begin{equation}\label{eq5}
\varphi \left( r \right) \approx {\varphi _0} + \left( {I\left( r \right) - {I_0}} \right){\left( {\frac{{\Delta I}}{{\Delta \varphi }}} \right)^{ - 1}},
\end{equation}
where ${\varphi _0} = {6^\circ}$ is the initial approximation to the magnet edge profile and $I\left( r \right)$ is the specified value of the integrated field strength obtained from Eqs.~\ref{eq:scaling_law} and~\ref{eq4}. The ratio $\frac{{\Delta I}}{{\Delta \varphi }}$ was obtained by creating another magnet model with ${\varphi _1} = {5^\circ}$ sector angle and subtracting the integrated field strengths obtained from both models. To simplify the task a discrete version of the good-field region of each magnet was considered. The radial coordinates ${r_1},\,{r_2},...,{r_5}$ split the good-field region into four intervals of equal length and the corresponding azimuthal coordinates $\varphi ({r_1}),\,\varphi ({r_2}),...,\varphi ({r_5})$ were obtained from Eq.~\ref{eq5}. The corrected pole edge shape was recovered by fitting a polynomial to the calculated nodes $\left\{ {{r_i},\,\varphi_{i}} \right\}$, $1\leq i\leq 5$. As Fig.~\ref{fig:fig5kiril} shows this procedure reduces the peak relative integrated field error by an order of magnitude. Further improvement can be obtained by ``tweaking'' the nodes individually until the desired accuracy goal has been reached. Figure~\ref{fig:fig6kiril} shows the pole edge profile of the F-magnet obtained from Eq.~\ref{eq5} compared to a $6^\circ$-edge and a $5^\circ$-edge.

\begin{figure}[htbp]
    \centering
    \begin{subfigure}[b]{0.48\columnwidth}
        \includegraphics[width=\columnwidth]{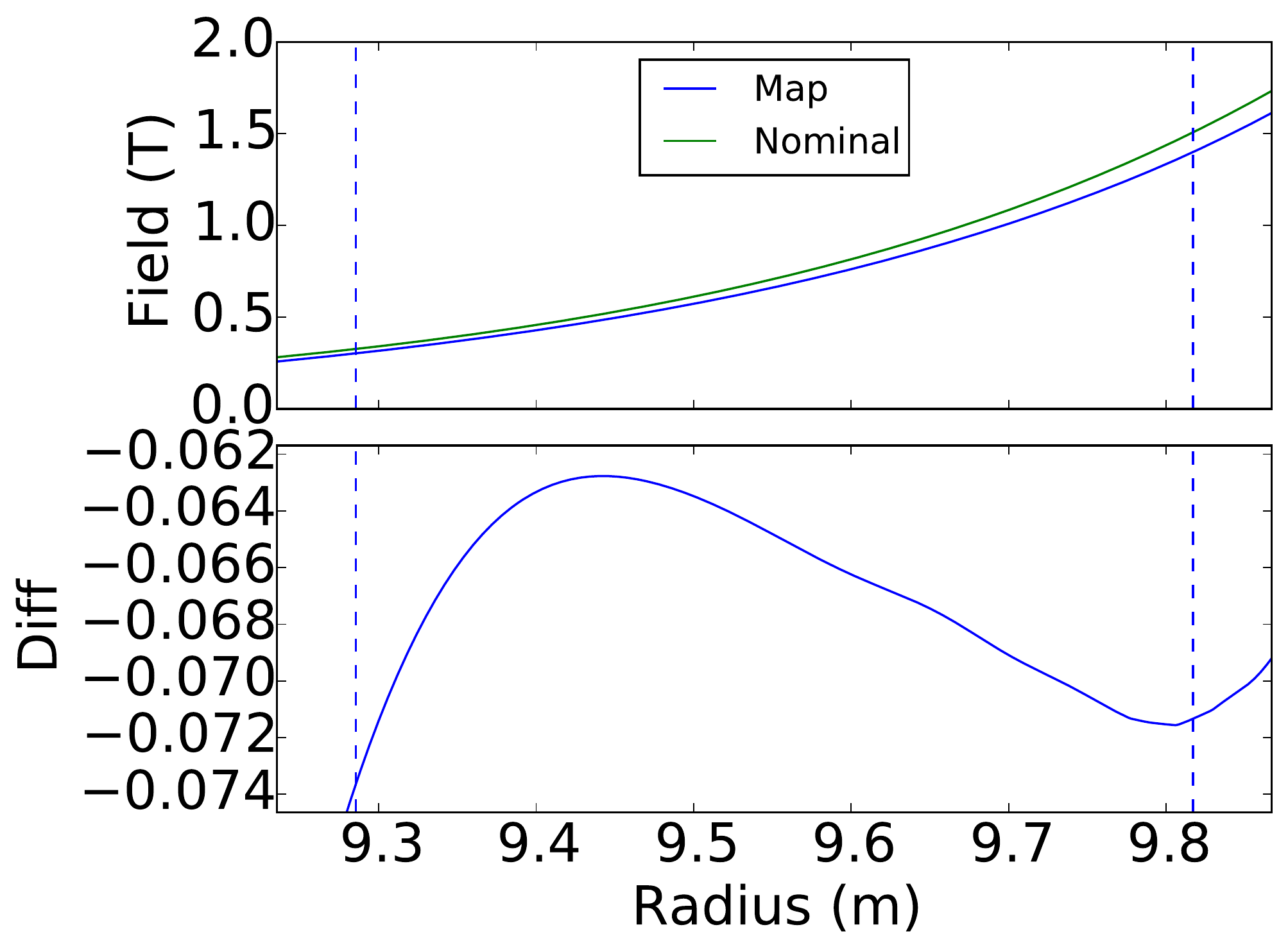}
        \caption{F peak field}
        \label{fig:norma_3d_F_midplane_pea}
    \end{subfigure}
    \begin{subfigure}[b]{0.48\columnwidth}
        \includegraphics[width=\columnwidth]{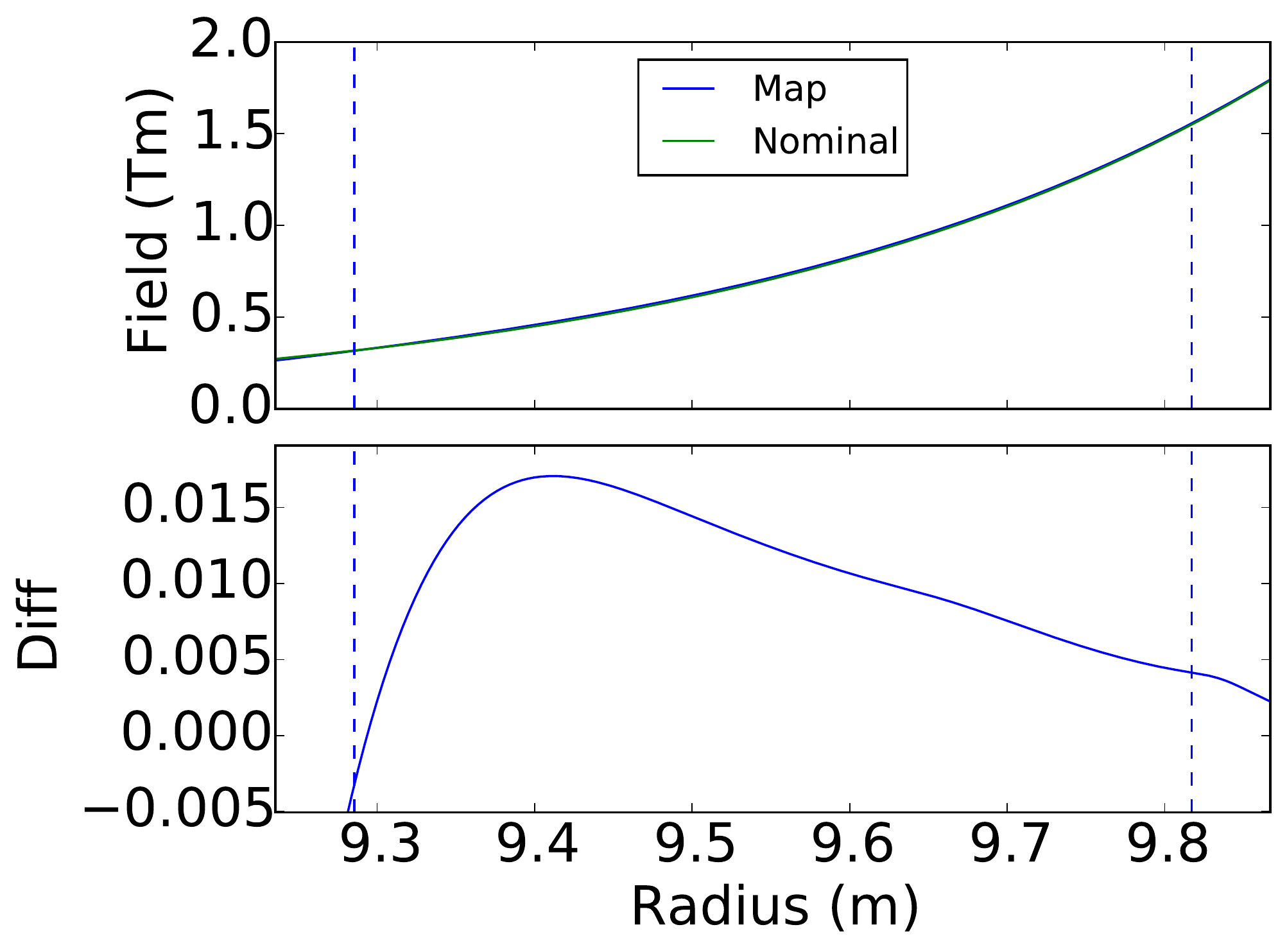}
        \caption{F integrated field}
        \label{fig:norma_3d_F_midplane_integrated}
    \end{subfigure}
    \begin{subfigure}[b]{0.48\columnwidth}
        \includegraphics[width=\columnwidth]{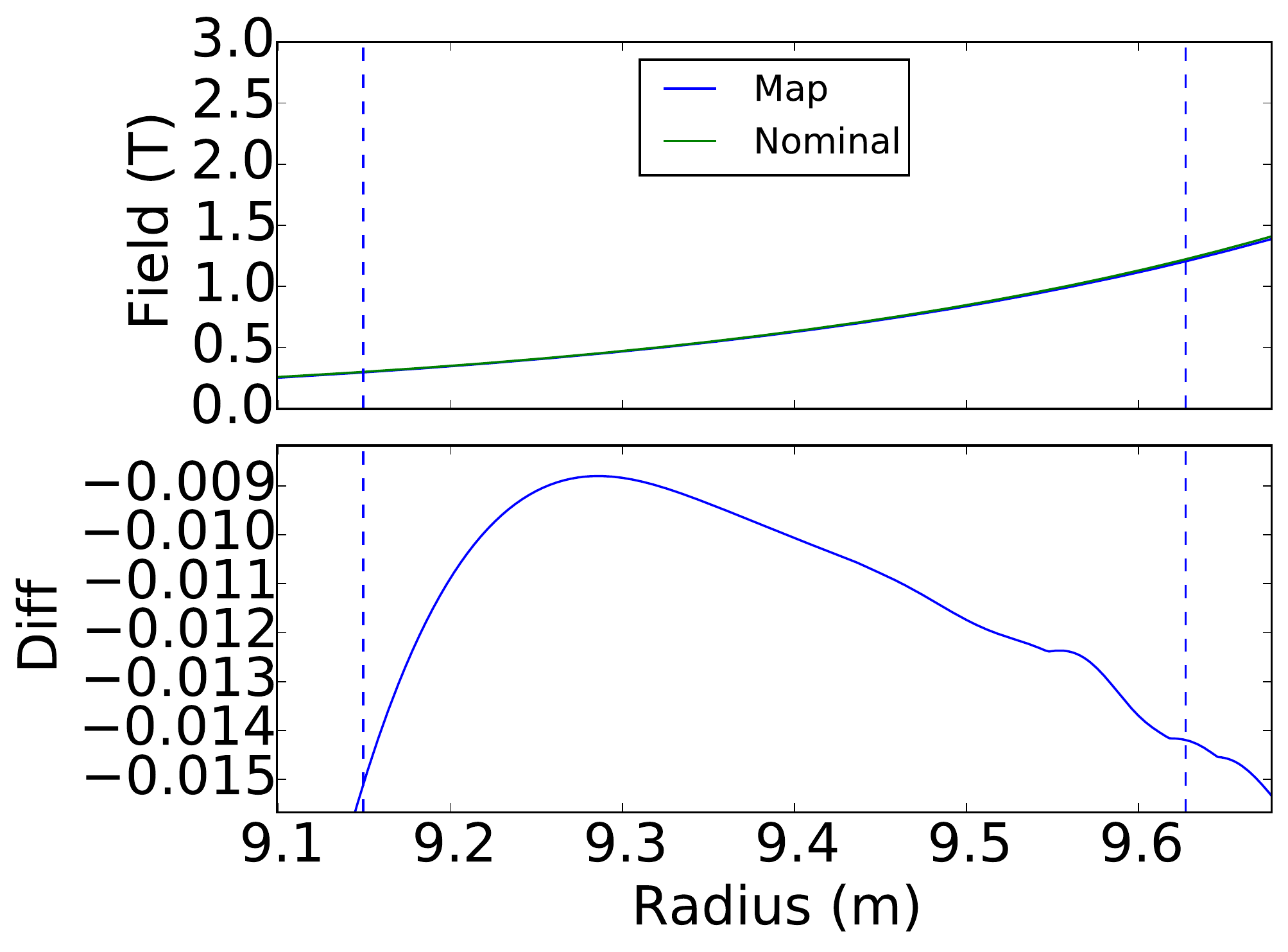}
        \caption{D peak field}
        \label{fig:norma_3d_D_midplane_pea}
    \end{subfigure}
    \begin{subfigure}[b]{0.48\columnwidth}
        \includegraphics[width=\columnwidth]{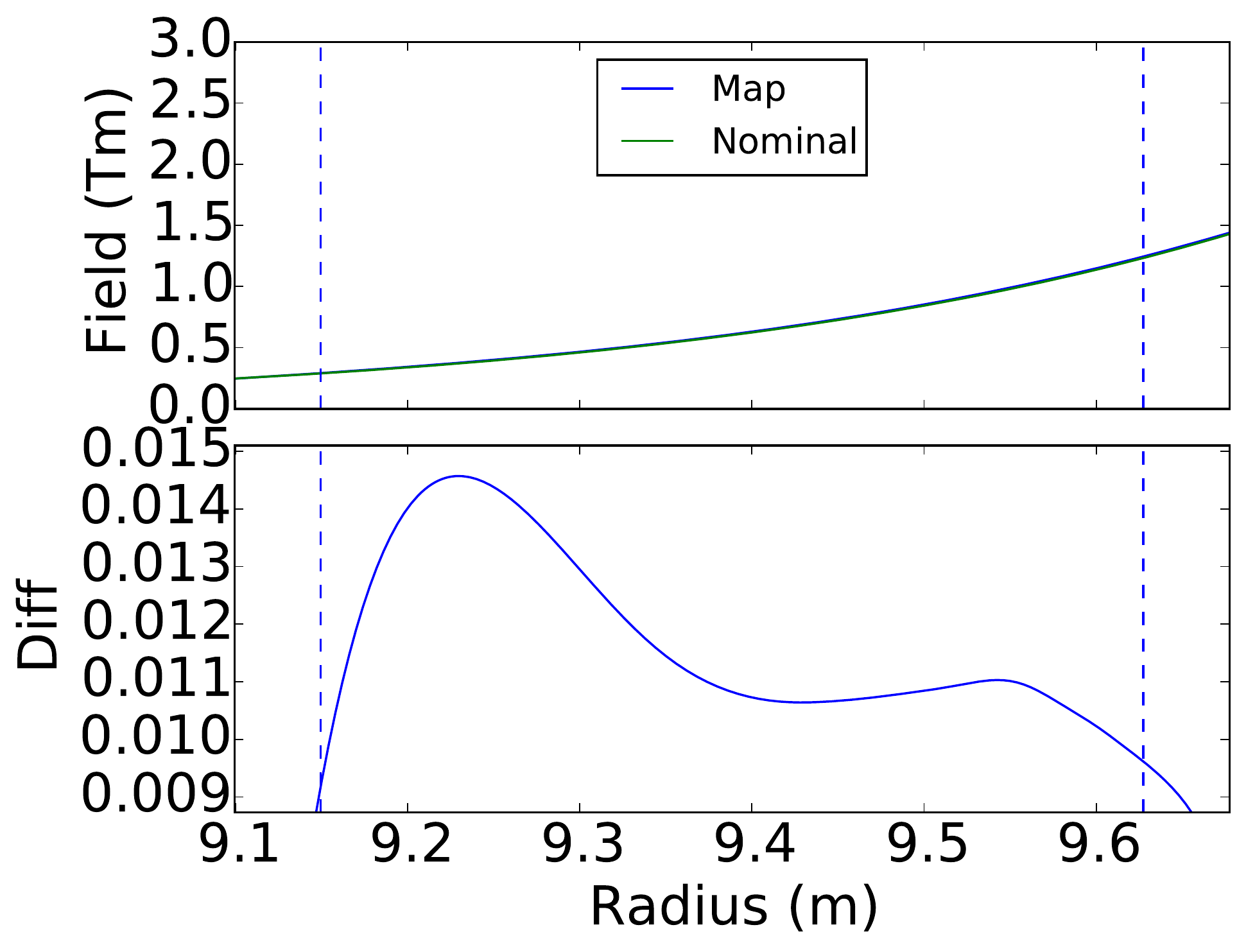}
        \caption{D integrated field}
        \label{fig:norma_3d_D_midplane_integrated}
    \end{subfigure}

    \caption{F and D magnet peak and integrated field for map compared to the nominal design. Dashed vertical lines show the good field region extents. The horizontal axis is the radial distance from the machine center.}
\label{fig:norma_3d_D_midplane_peak_int}
\end{figure}

Whilst the radial profile in the final 3D magnets deviates from that of the nominal design, the integrated field along each arc of constant radius is well matched by this adjustment of the edge geometry. Figure~\ref{fig:norma_3d_D_midplane_peak_int} show the peak and integrated fields for the focusing and defocusing magnets. The peak field is different from the nominal values by up to 7.3~\% in the F magnet and 1.5~\% in the D magnet, whereas the greatest deviation from nominal for the integrated field is 1.7~\% in the F and 1.5~\% in the D.

The 2D and 3D magnet models can now be used to obtain field maps for particle tracking. In following sections we analyze the beam dynamics with these realistic magnets models.

\section{Beam dynamics with 2D magnets}
\label{sec:2d_magnets}

In this section we model NORMA using the field profiles from the 2D magnet models, to compare the dynamics with those due to the nominal fields.

\subsection{Implementing the 2D field maps in Zgoubi}

The 2D magnet design is output from OPERA as a table of the vertical component of the field, $B_y$, along a radial line through the magnet. The $B_x$ and $B_z$ are zero by symmetry.
These radial profiles can be used in Zgoubi with either the \texttt{POLARMESH} or \texttt{DIPOLES} elements.

\begin{figure}[htbp]
\centering
\includegraphics{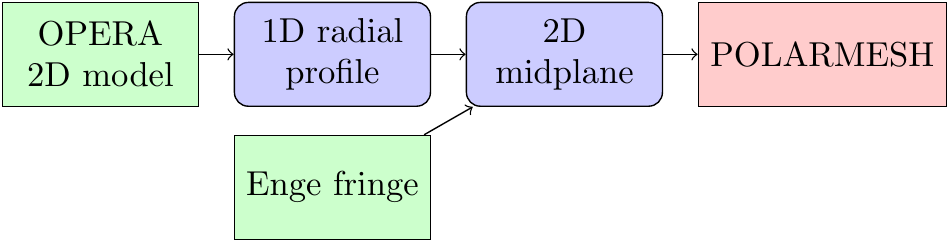}
\caption{Using the 2D OPERA model as input to the POLARMESH element.}
\label{fig:2d_polarmesh}
\end{figure}

To track the 2D model with \texttt{POLARMESH} a midplane field map must be generated. For each magnet the radial profile is combined with a longitudinal Enge fringe function (Eq.~\ref{eq:enge_fringe}) to give the field at each point on a 2D mesh. We use the same Enge parameters as in the nominal model. These are summed to give the whole cell midplane field map. The process is shown in Fig.~\ref{fig:2d_polarmesh}.

\tikzstyle{inb} = [rectangle, draw, fill=green!20,text width=5em, text centered, minimum height=3em, node distance=1em]
\tikzstyle{proc} = [rectangle, draw, fill=blue!20,text width=5em, text centered, rounded corners, minimum height=3em, node distance=1em]
\tikzstyle{outb} = [rectangle, draw, fill=red!20,text width=6.5em, text centered, minimum height=3em, node distance=1em]

\begin{figure}[htbp]
\centering
\includegraphics{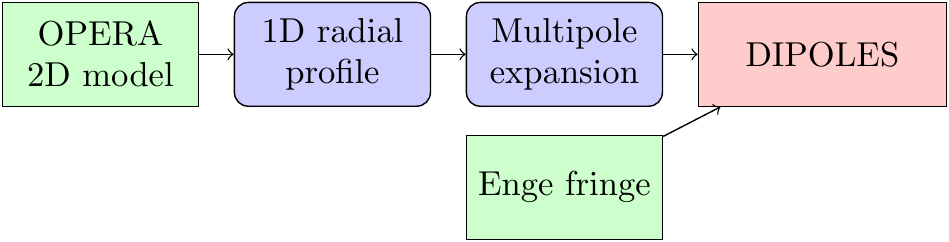}
\caption{Using the 2D OPERA model as input to the DIPOLES element.}
\label{fig:2d_dipoles}
\end{figure}

To use the \texttt{DIPOLES} element, a set of multipole coefficients are found by fitting a polynomial to the radial profile. These coefficients are used along with parameters of the fringe fields and geometry, as shown in Fig.~\ref{fig:2d_dipoles}.

The multipole fit for \texttt{DIPOLES} can be a global fit across the whole magnet giving a single model that can be used at every energy. Alternatively, a region around each orbit can be selected and fitted. The latter gives a higher accuracy, avoiding any smoothing from the fit, but means that the magnet description must be refitted at every energy. For simulations, we used \nth{9}-order local fits about each energy.

\subsection{Dynamics and DA with 2D magnets}
\label{sec:dynamics_with_2D_magnets}

\begin{figure}[htbp]
  \centering
  \includegraphics[width=0.9\columnwidth]{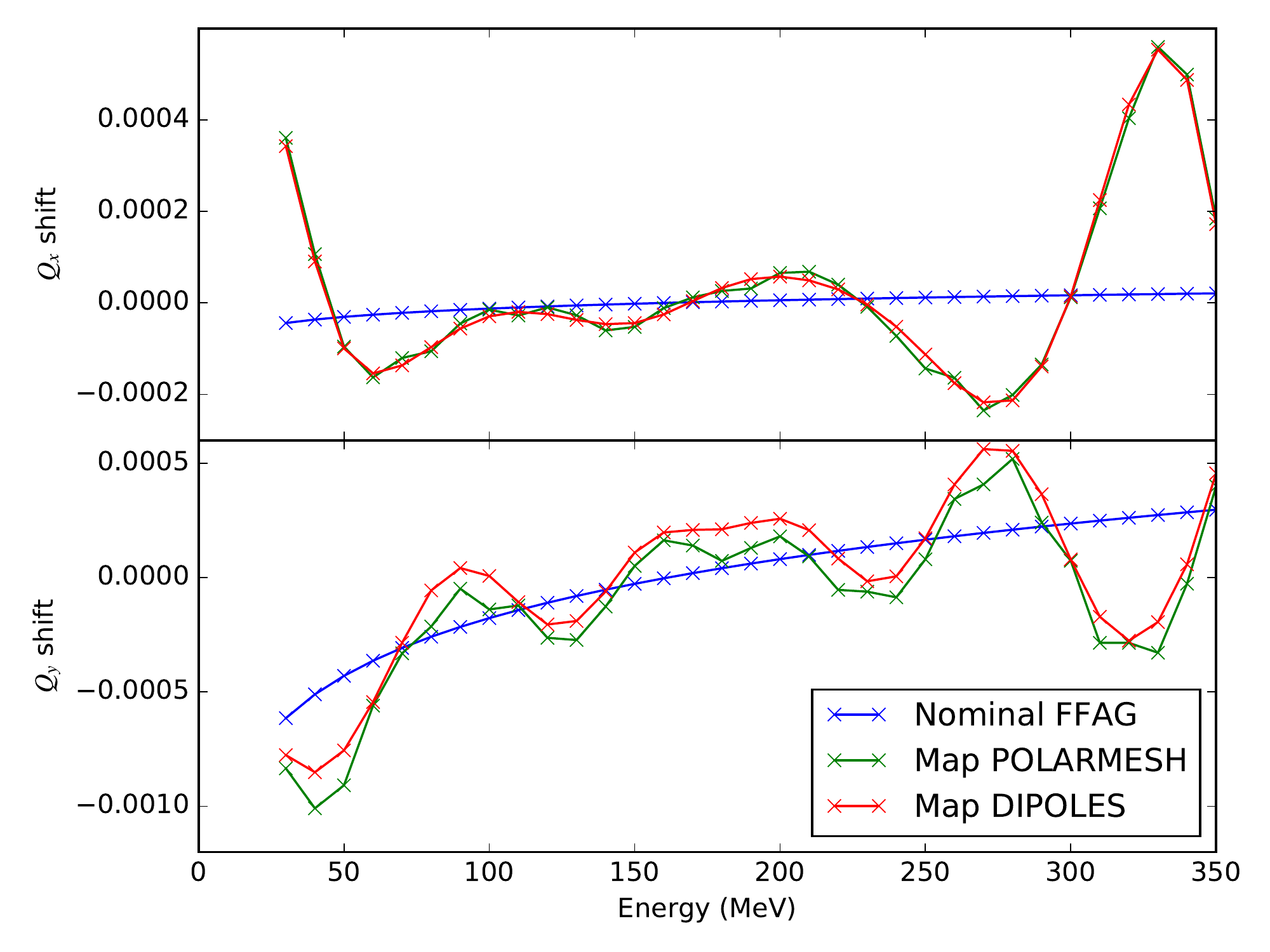}
  \caption{Cell tune shift for 2D magnets modelled with POLARMESH and DIPOLES compared to the nominal design.}
\label{fig:norma_2d_da_energy_tune}
\end{figure}

The 2D model gives good agreement with the nominal design. Figure~\ref{fig:norma_2d_da_energy_tune} shows the horizontal and vertical tunes for the 2D models against the nominal design. The mean tune is within \SI{e-4} in both planes and the tune excursion grows from \SI{6.4e-5} and \SI{9.1e-4} to \SI{7.9e-4} and \SI{1.5e-3} between the nominal and the \texttt{POLARMESH} model. Agreement between the \texttt{POLARMESH} and \texttt{DIPOLES} simulations methods are very good, demonstrating that the multipole expansion is an effective method to use for tracking.

\begin{figure}[htbp]
  \centering
  \includegraphics[width=0.9\columnwidth]{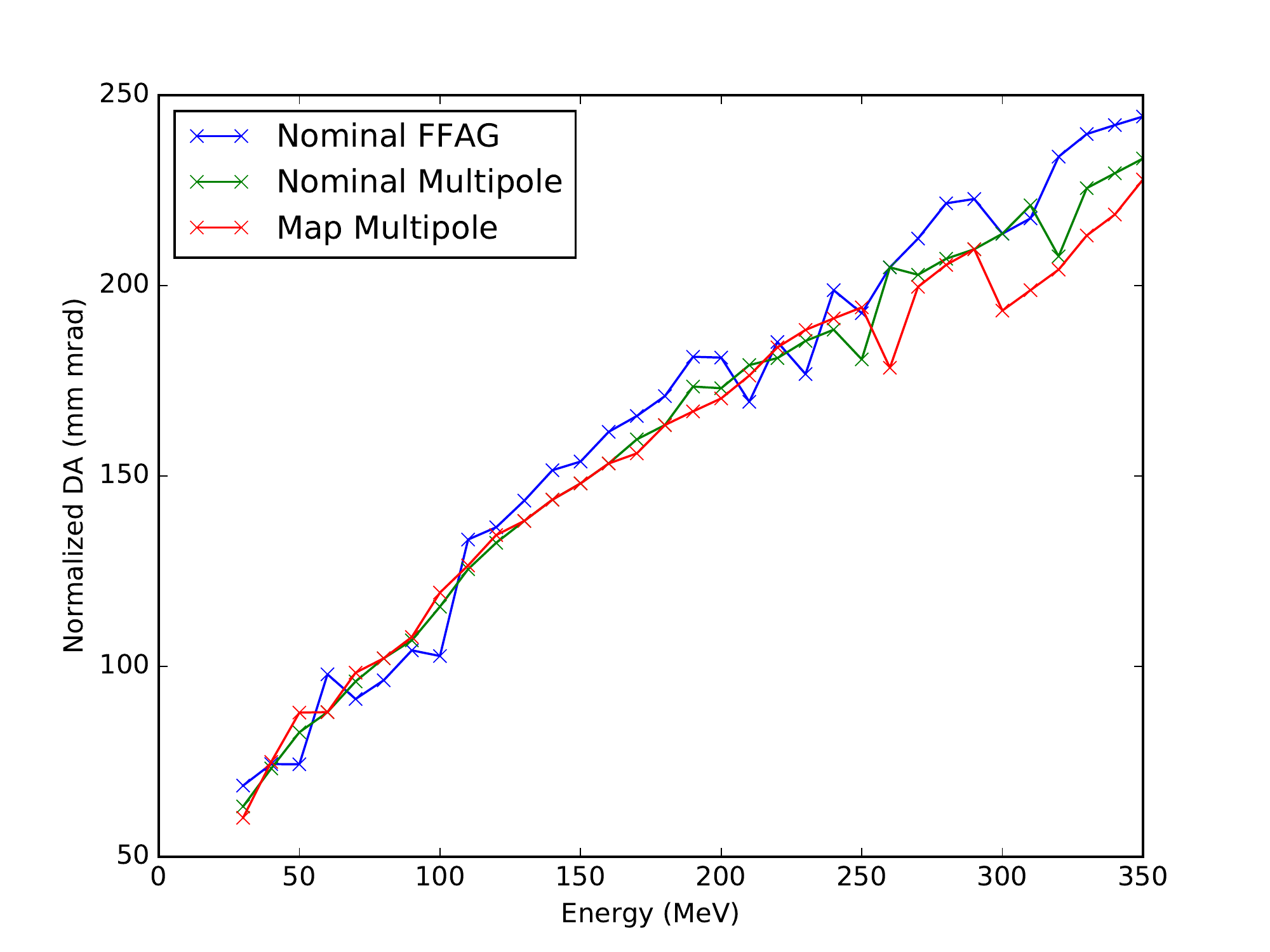}
  \caption{DA for 2D field profiles compared to nominal magnets.}
\label{fig:norma_da_energy_2d_dipoles}
\end{figure}

An increased tune excursion may cause accelerated particles to cross resonances that reduce their stability. In order to confirm that this is not the case for the 2D magnets we determine the 1000-turn DA. Using \texttt{DIPOLES} with fits to the 2D radial field profile at each orbit, we find a 45\degree~ DA (i.e. where particle amplitude is increased equally in both transverse planes) that is very close to the nominal lattice over the full range of energies; this is shown in Fig.~\ref{fig:norma_da_energy_2d_dipoles}. The figure shows that deviations from the ideal field in the 2D model do not cause a significant drop in DA. As with the nominal design, the DA is kept above 50~mm\,mrad over the full range of energies in the accelerator.

\begin{figure}[htbp]
    \centering
        \includegraphics[width=\columnwidth]{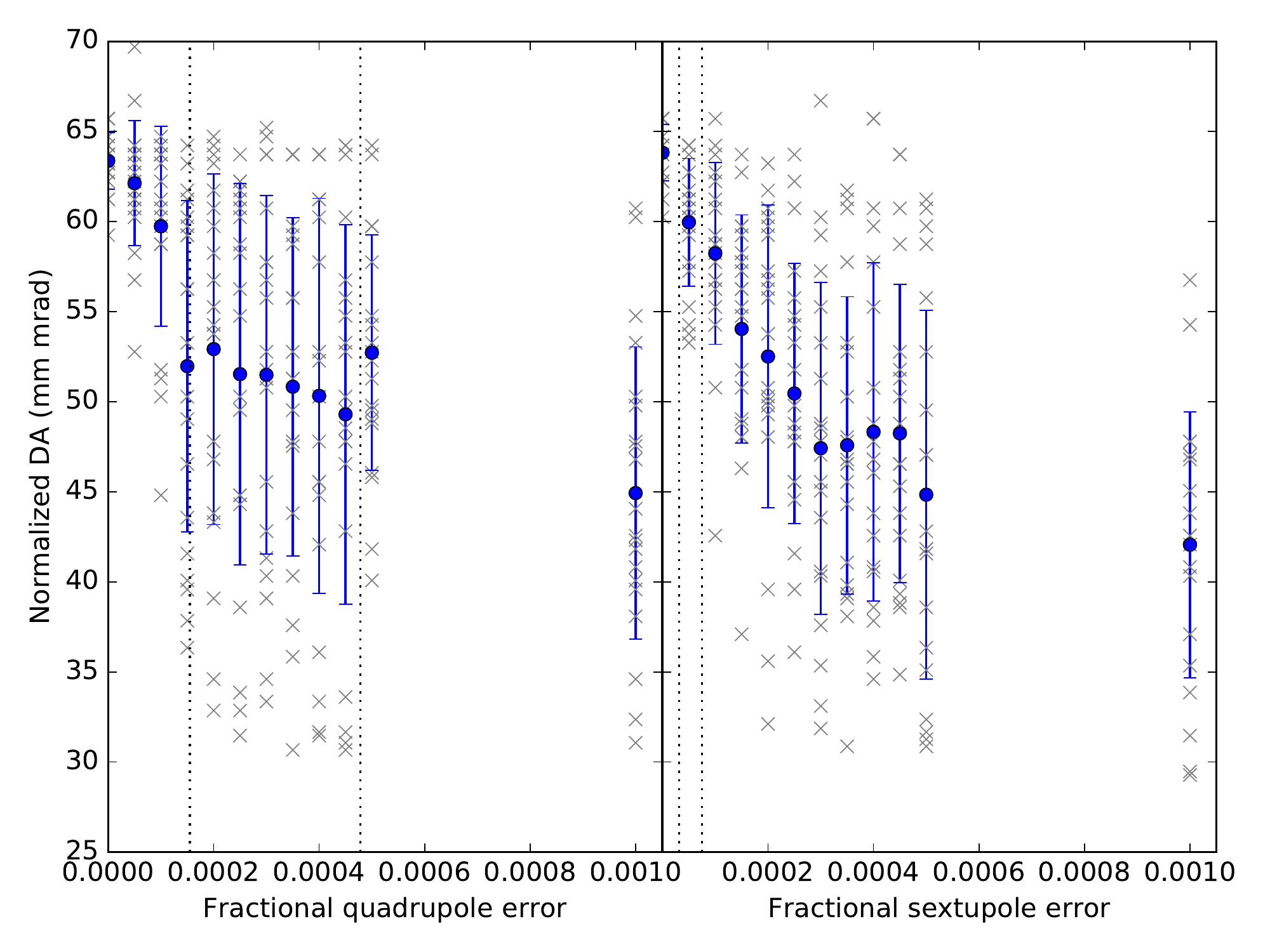}
        \label{fig:da_30_quad_errors}
    \caption{DA for random quadrupole and sextupole error distributions around the 30~MeV orbit. Crosses show individual seeds and the blue bars show their mean and standard deviation. Dotted vertical lines represent the actual size of the errors in the 2D magnet model.}
\label{fig:da_30_errors}
\end{figure}

We can also add random multipole errors to the magnets. By using errors of a similar magnitude to those seen in the 2D field we can obtain some understanding of how the worst-case errors might affect the DA. However, applied multipole errors tend to cause big field changes away from the point that they are applied, which would not be seen in a well-designed real magnet. For a given error size, the ideal field is expanded around the mean position of a given orbit; Gaussian-distributed errors are then applied to a given multipole component. Figure~\ref{fig:da_30_errors} shows how the DA is reduced as quadrupole and sextupole errors applied around the 30~MeV orbit increase, using 20 seeds per error size; the dotted vertical lines represent the actual size of the errors in the 2D magnet model. We can see that quadrupole and sextupole errors of magnitudes similar to those in the 2D OPERA model do not cause the DA to drop significantly below 50~mm\,mrad.

Overall, the deviations of the 2D profile from the nominal design do not affect the dynamics enough to cause problems. We find that the increased tune shifts as a function of energy are tolerable. Random quadrupole and sextupole errors of similar magnitudes can also be added without causing the DA to drop significantly below our 50~mm\,mrad requirements.

Whilst the 2D simulations are a good indicator that it is feasible to produce the magnetic field profile specified in the nominal lattice design, the following sections show that the longitudinal and full 3D fields are needed for a sufficient understanding of the lattice dynamics. 2D modelling is however a useful step in the design and optimization stage, as it can be performed more rapidly than with full 3D models.

\section{Beam dynamics with fringe fields}
\label{sec:fringe_fields}

In this section the effect of the fringe extent and radial dependence is investigated. For the initial lattice design an Enge fringe field was used, with a 4~cm extent constant with respect to radius. In the real magnet, the fringe extent depends on the pole shapes. It should be expected that adjustments to the magnet strengths and/or field profile will be needed to account for the realistic fringe fields.

The fringe-field extent has a significant effect on the edge focusing of the magnets and therefore contributes to the tune of the cell.
Changes in the fringe-field extent as a function of radius will therefore have an effect on the tune as a function of energy.

\begin{figure}[htbp]
    \centering
        \includegraphics[width=\columnwidth]{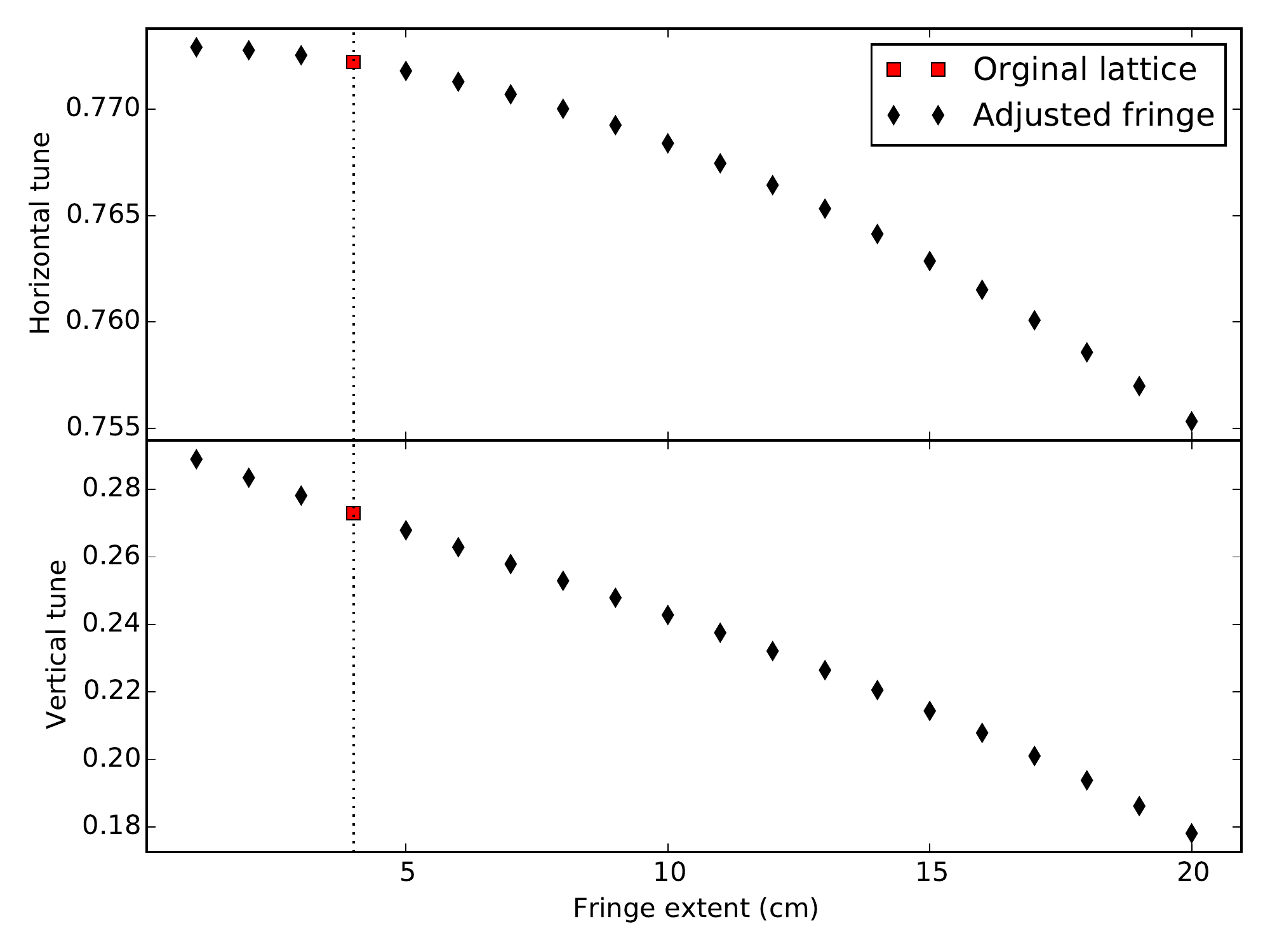}
    \caption{Effect on cell tune at 30~MeV as the fringe field extent length is varied.}
\label{fig:fringe_sensitivity_NU}
\end{figure}

\begin{figure}[htbp]
  \centering
  \includegraphics[width=0.9\columnwidth]{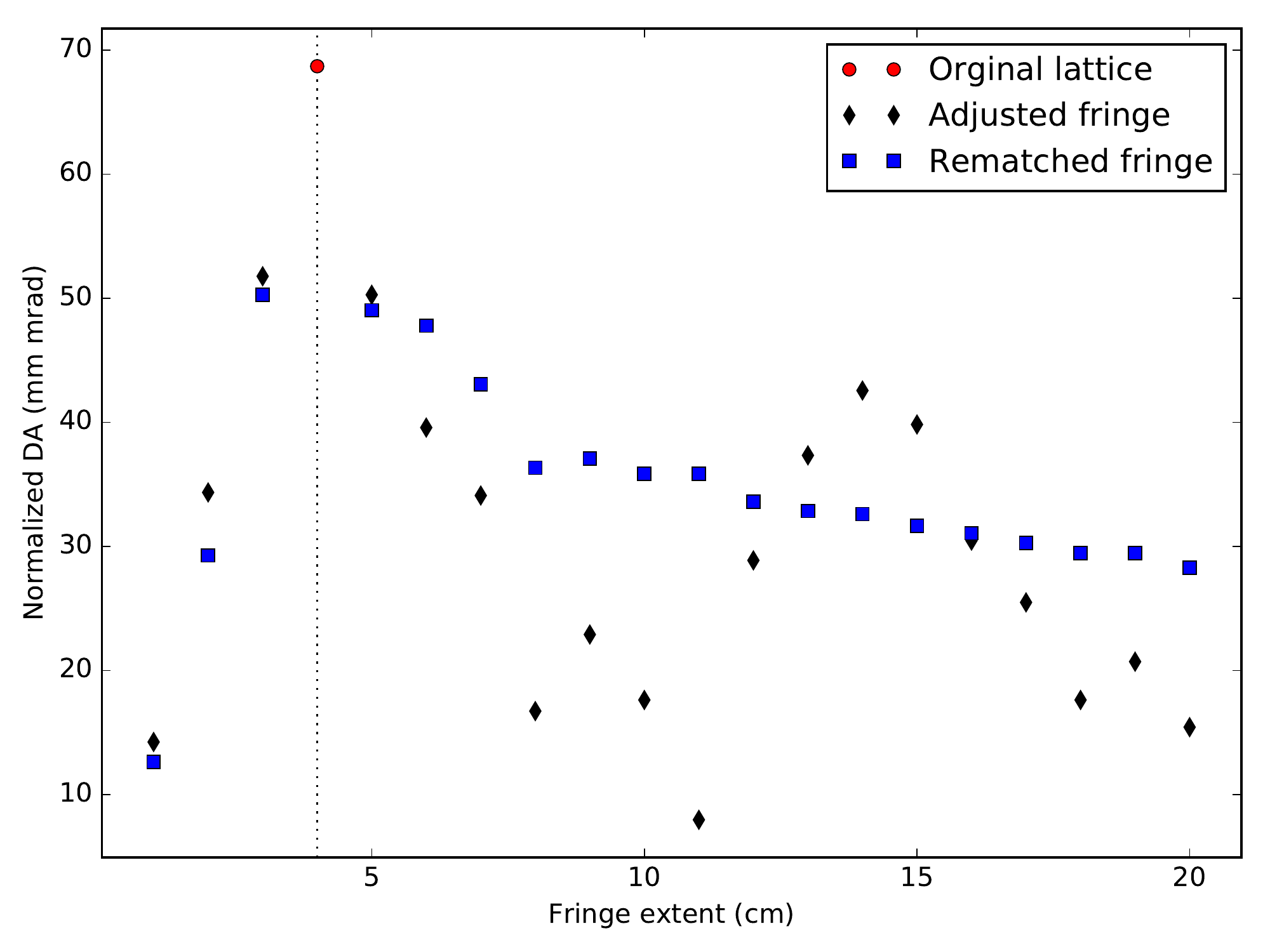}
  \caption{Black diamonds show the DA at 30~MeV as the fringe length is adjusted. Recovery of DA by rematching tune to the original value is shown in blue squares.}
\label{fig:fringe_sensitivity_DA_kappa0_rematch}
\end{figure}

To show the effect on the tune, the fringe extent was varied while all other lattice and magnet parameters were held fixed and no rematching was performed. Figure~\ref{fig:fringe_sensitivity_NU} shows the change in tunes as a function of fringe extent; the original value of 4~cm in the nominal lattice is highlighted. As the tune shifts, the working point moves and approaches resonances. At an extent of around 10~cm the vertical tune approaches 0.25; the black points in Fig.~\ref{fig:fringe_sensitivity_DA_kappa0_rematch} show how this causes a large drop in DA as the working point approaches a fourth-order resonance.

\begin{figure}[htbp]
    \centering
        \includegraphics[width=\columnwidth]{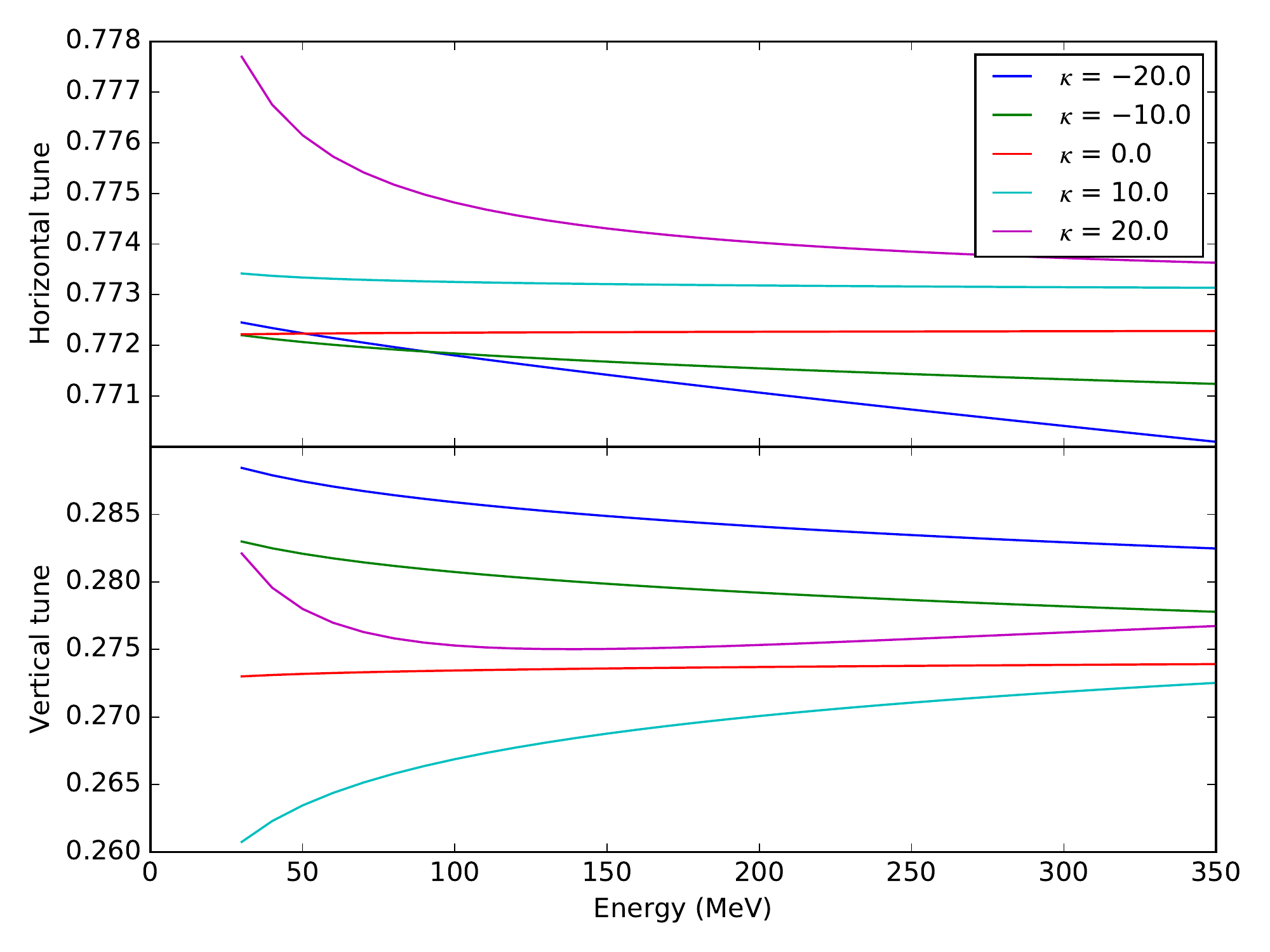}
        \label{fig:fringe_compare_NU_Y_kappa}
    \caption{Effect on cell tune of fringe field that varies with radius.}
\label{fig:fringe_compare_NU_kappa}
\end{figure}

The fringe-field extent can be varied by extending the fringe-field description from Eq.~\ref{eq:enge_fringe} with the variable $\kappa$ and making $\lambda$ a function of $r$, so that
\begin{equation} \label{eq:ff_kappa}
\lambda = \lambda_0 (r_0 / r)^\kappa .
\end{equation}
A positive $\kappa$ gives a fringe field with a larger extent at smaller radii, as would be expected due to the larger pole gap. Figure~\ref{fig:fringe_compare_NU_kappa} shows the tune as a function of energy for a range of $\kappa$ values. These tune shifts with energy can cause problems as the beam will cross resonances during acceleration, so it is important that they can be corrected.

The tune shift due to a change in the fringe field can be compensated by changing the magnet body fields. For a given fringe extent the working point of the lattice can be rematched to recover the original tunes by adjusting the magnet strengths $B_0$ of the F and D magnets, and the field index $k$ shared by both F and D magnets. The match is constrained to keep the outer closed-orbit position constant. DA is most critical at injection energy before the emittance is reduced somewhat by adiabatic damping, so the rematching was performed at 30~MeV. PyZgoubi's optimization feature, making use of the Nelder-Mead downhill simplex method, was used for the rematching.
Figure~\ref{fig:fringe_sensitivity_DA_kappa0_rematch} shows how the rematch recovers some of the lost DA for lattices where the change in fringe field caused the tune to approach a resonance.

\begin{figure}[htbp]
  \centering
  \includegraphics[width=0.9\columnwidth]{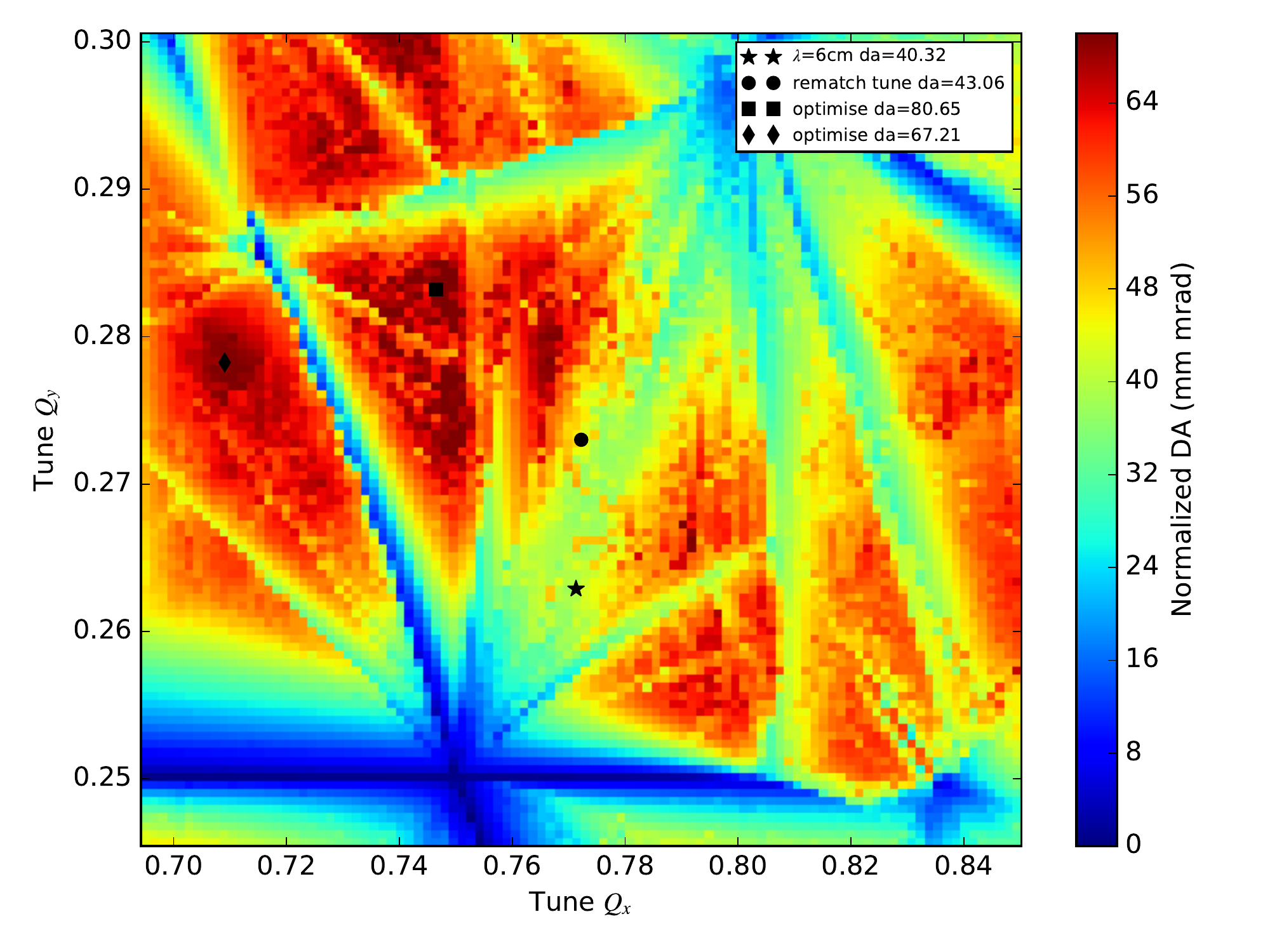}
  \caption{Rematching the cell tunes with a 6~cm fringe extent at 30~MeV. The star marker shows the working point shift due to changing the fringe extent. The dot shows the working point after rematching to the original tunes. The square and diamond markers show the highest DA point, and the region of highest DA.}
\label{fig:norma_fringe_rematch_scan_6}
\end{figure}

We find that the original working point is not necessarily optimal for maximising DA, as the shape of the fringe field can affect the relative strength of the resonances. It is therefore important to re-optimize the working point to find the largest DA region. Re-optimization is performed by scanning the lattice through a range of working points and calculating the dynamic aperture at each point. Figure~\ref{fig:norma_fringe_rematch_scan_6} shows how the working point is rematched for a fringe extent of 6~cm. Rematching the tune to the original values recovers some DA, but the scan is required to recover a DA above 50~mm\,mrad.

The extent and shape of the fringe fields play a large role in the dynamics of an accelerator. The lattice must be re-optimized to account for their effects. This shows that 3D magnet simulation, which will give the fringe field, is a necessary step for realistic lattice design.

\section{Beam dynamics with 3D magnets}
\label{sec:3d_magnets}

In this section we model NORMA using field maps from 3D magnet simulations. These contain not just the radial field profile but also how the field falls off outside the magnet body.

\subsection{Implementing the 3D field maps in Zgoubi}

\begin{figure}[htbp]
    \centering
    \begin{subfigure}[b]{0.49\columnwidth}
        \includegraphics[width=\textwidth]{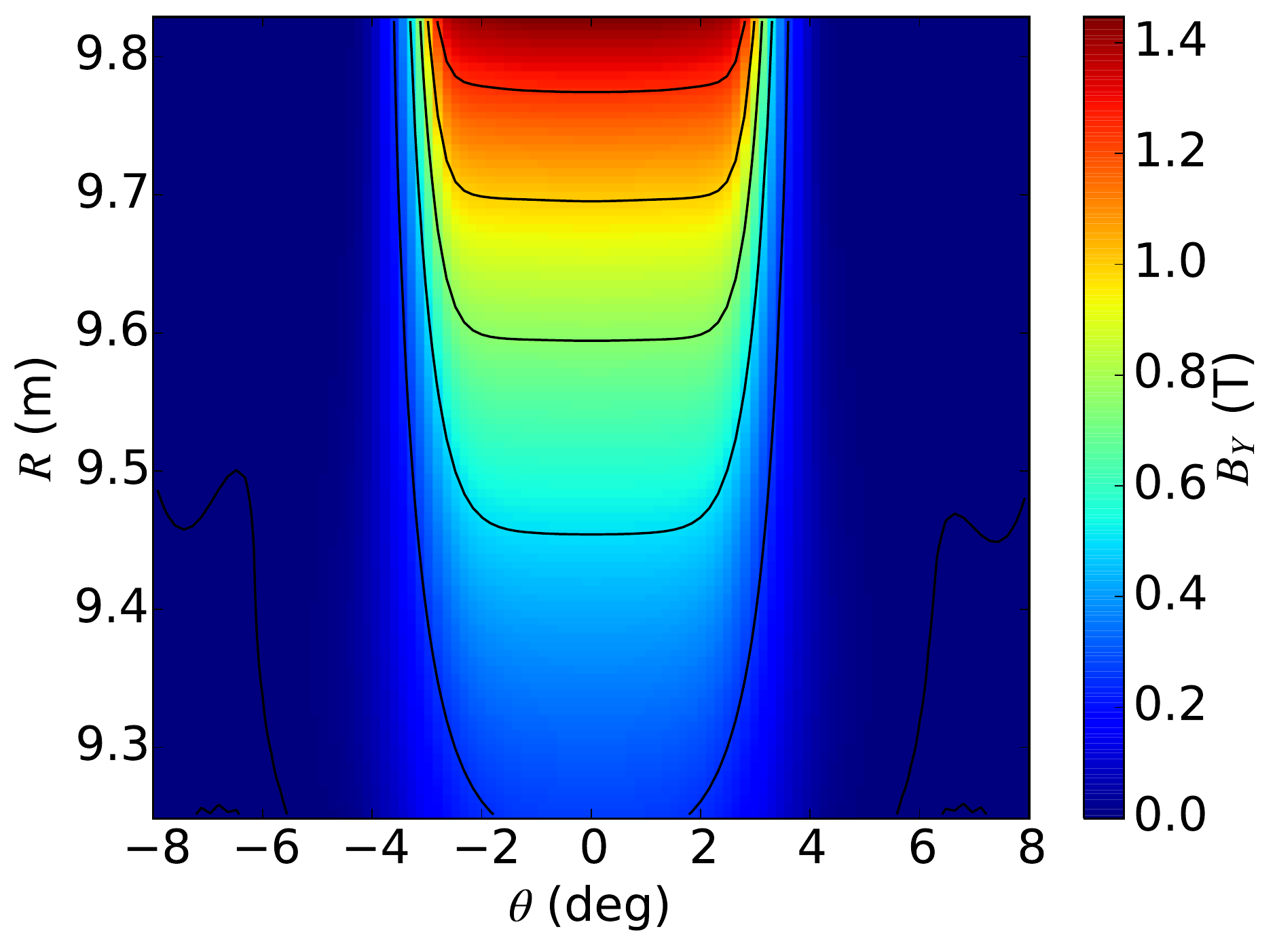}
        \caption{F midplane field}
        \label{fig:3d_F_midplane}
    \end{subfigure}
    \begin{subfigure}[b]{0.49\columnwidth}
        \includegraphics[width=\textwidth]{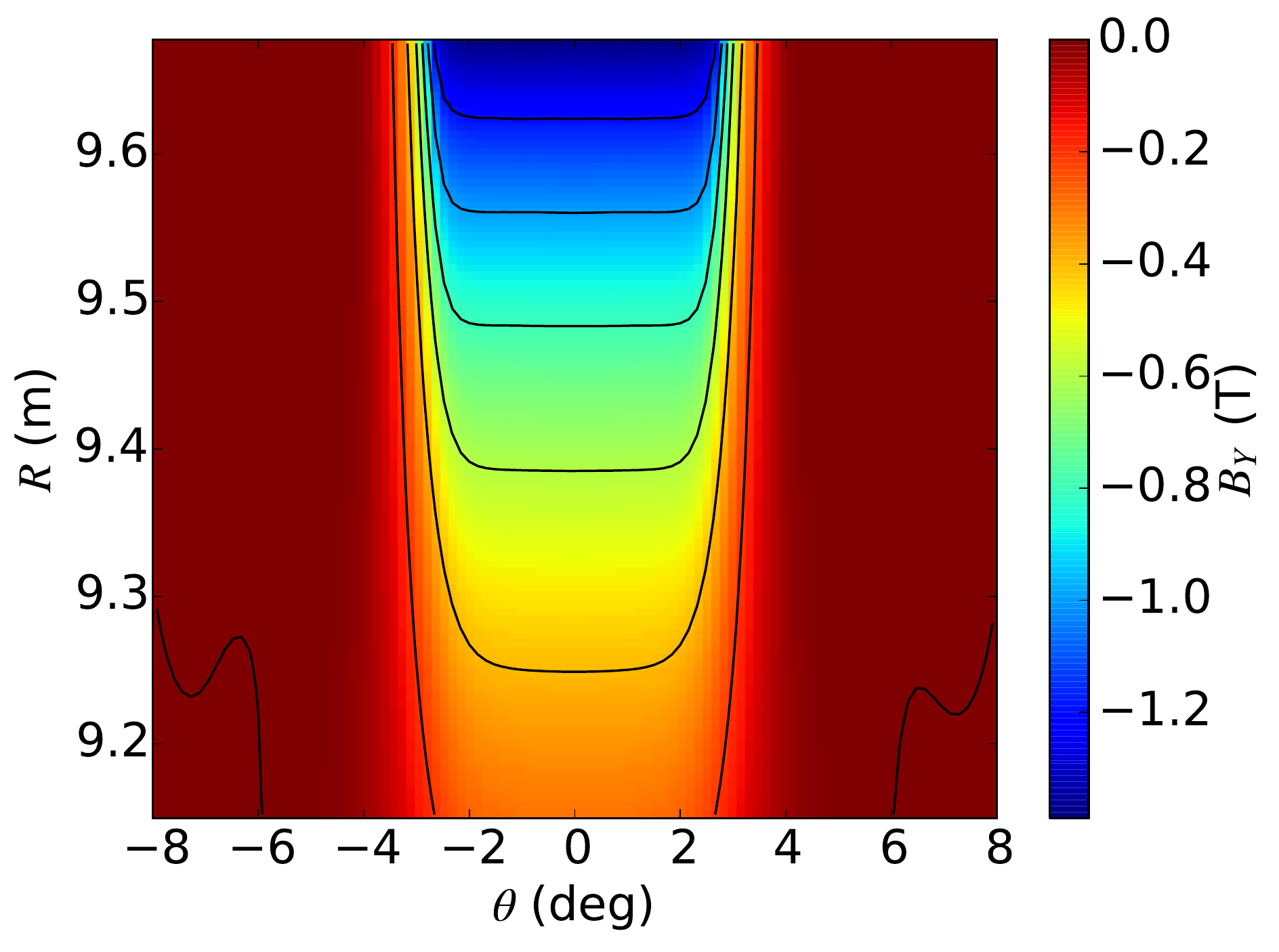}
        \caption{D midplane field}
        \label{fig:3d_D_midplane}
    \end{subfigure}

    \caption{Full midplane fields from 3D OPERA simulation.}
\label{fig:3d_midplane}
\end{figure}

3D magnets are output from the OPERA simulation as midplane field maps, i.e. the $B_y$ component at grid points on a horizontal midplane; the $B_x$ and $B_z$ are zero by symmetry. The midplane is sufficient to fully define the field over the vacuum region of the magnet including the fringe field, while significantly reducing the computational resources required to generate and store the field map. Figure~\ref{fig:3d_midplane} shows the midplane fields for each magnet.

The dynamics in the full magnet design will be strongly influenced by both the body field and the fringe fields. For example, the reduced peak field in the F magnet is compensated by an extended field length such that the integrated field is close to the nominal design. The 3D field maps must therefore be used as a whole, as the radial and fringe parts cannot be independently combined with the nominal field.

\begin{figure}[htbp!]
\centering
\includegraphics{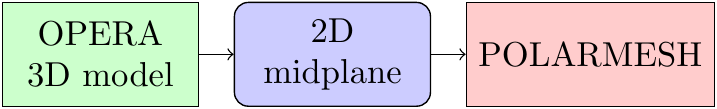}
\caption{Using the 3D OPERA model as input to the POLARMESH element.}
\label{fig:3d_polarmesh}
\end{figure}

\begin{figure}[htbp]
  \centering
  \includegraphics[width=0.9\columnwidth]{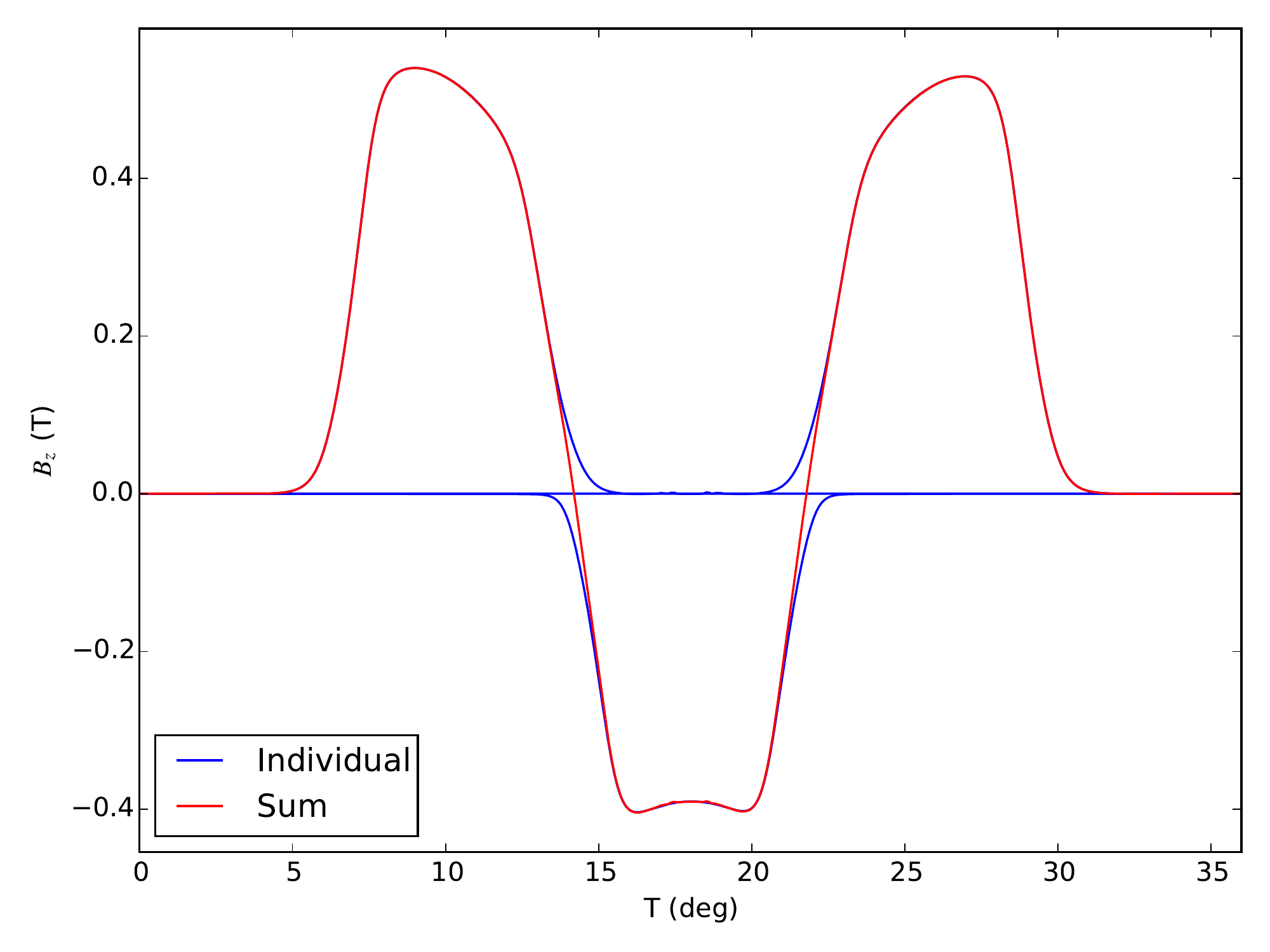}
  \caption{Overlap of individual magnet fields and the full field found by linear summation, along the 50~MeV orbit.}
\label{fig:overlap_50}
\end{figure}

The midplane maps from the 3D magnet simulation can be used in Zgoubi with either the \texttt{POLARMESH} or \texttt{DIPOLES} elements. The F and D magnets are designed independently, as this allows simple boundary conditions and reduction in computation time due to symmetry. The midplane maps for the magnets are combined assuming a linear superposition to create the map for the full cell, which is then read into the \texttt{POLARMESH} element as shown in Fig.~\ref{fig:3d_polarmesh}. At the field crossover point in the overlap between the F and the D magnets, the residual field from the magnets is about 5-10\% of the body field, greatest at the low-energy orbits. Figure~\ref{fig:overlap_50} shows the overlap at 50~MeV. A further step would be to combine the magnets within the FEM simulations so that the overlap region is fully calculated. If this is found to cause a significant effect, then either clamp plates can be used to reduce the fringe extent or rematching as described below can be used to account for it.

\begin{figure}[htbp!]
\centering
\includegraphics[width=0.99\columnwidth]{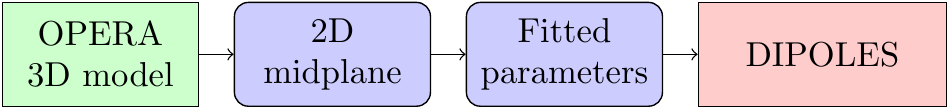}
\caption{Using the 3D OPERA model as input to the DIPOLES element.}
\label{fig:3d_dipoles}
\end{figure}

Alternatively, the 3D magnet can be modelled with a \texttt{DIPOLES} element, as shown in Fig.~\ref{fig:3d_dipoles}. This makes rematching simpler by reducing the description to a smaller set of variables, gives more reliable long-term tracking and improves performance compared to tracking in the map directly.
The \texttt{DIPOLES} element has several parameters that allow it to model a range of magnets. The radial profile is specified as a 9th-order polynomial, the entrance and exit boundaries can be moved and rotated independently, the fringe can be specified with up to 6 Enge coefficients and the fringe extent can be a function of radius.

We fit the \texttt{DIPOLES} using only the magnetic field and then use tracking to verify that the solution is good. For the initial fit we minimised the difference between the map and the \texttt{DIPOLES} at each grid point with in the good field region, using a Nelder-Mead downhill simplex, allowing the above parameters to vary. This however did not give a good match for the dynamics. Much better dynamic agreement was found by including the differences of the integrated field and integrated gradient along the particle trajectories in the objective function.

The dynamics of the resulting 3D fit are shown, compared to the \texttt{POLARMESH} and the nominal FFAG in Figs. \ref{fig:3d_norma_da_energy_NU_Y} and \ref{fig:3d_norma_da_energy_NU_Z}.

\tikzstyle{inb} = [rectangle, draw, fill=green!20,text width=5em, text centered, minimum height=3em, node distance=1em]
\tikzstyle{proc} = [rectangle, draw, fill=blue!20,text width=5em, text centered, rounded corners, minimum height=3em, node distance=1em]
\tikzstyle{outb} = [rectangle, draw, fill=red!20,text width=6.5em, text centered, minimum height=3em, node distance=1em]

\subsection{Dynamics with 3D magnets}
\label{sec:dynamics_with_3D_magnets}

\begin{figure}[htbp]
  \centering
  \includegraphics[width=0.9\columnwidth]{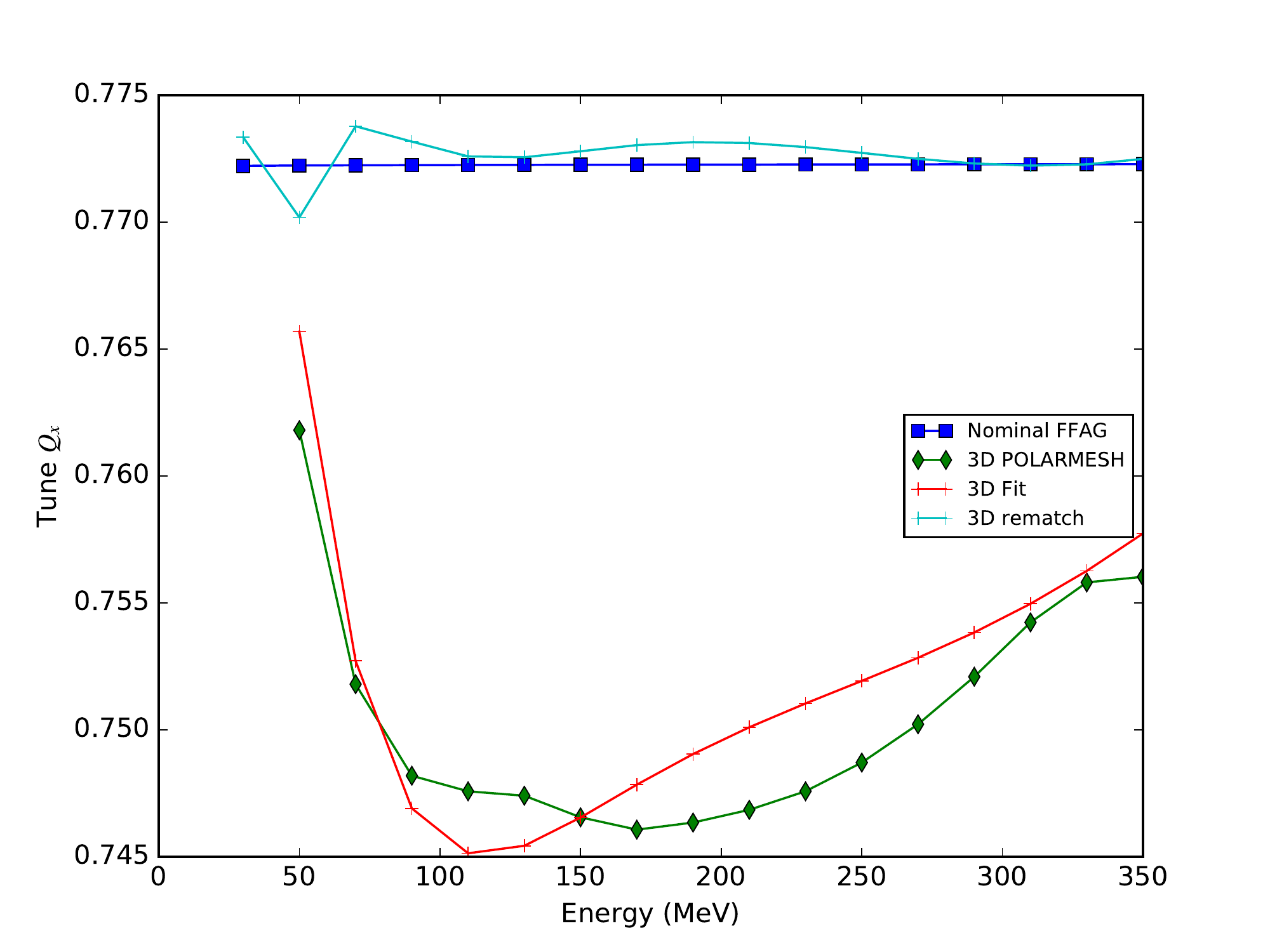}
  \caption{Horizontal cell tune as a function of energy for 3D magnets and rematched fields.}
\label{fig:3d_norma_da_energy_NU_Y}
\end{figure}

\begin{figure}[htbp]
  \centering
  \includegraphics[width=0.9\columnwidth]{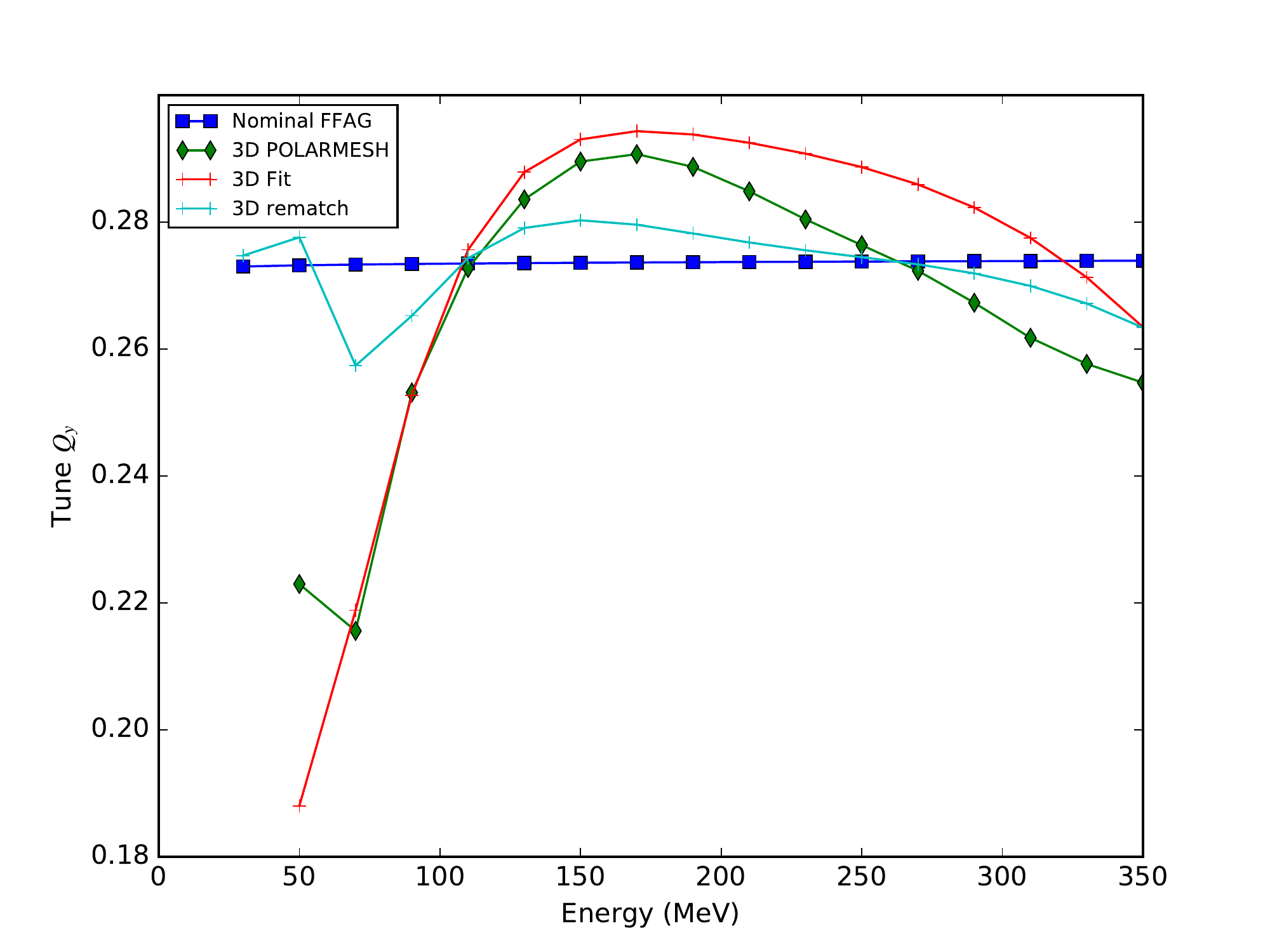}
  \caption{Vertical cell tune as a function of energy for 3D magnets and rematched fields.}
\label{fig:3d_norma_da_energy_NU_Z}
\end{figure}

The difference between the nominal fields and the 3D maps cause significant changes to the tune of the lattice. Figures~\ref{fig:3d_norma_da_energy_NU_Y} and~\ref{fig:3d_norma_da_energy_NU_Z} show the tune for \texttt{POLARMESH} in green and fitted \texttt{DIPOLES} in red, compared to the nominal \texttt{FFAG} in blue. It can be seen that the fit does a good job of reproducing the dynamics of the field map. At low energies the vertical tune with the 3D magnets crosses the quarter-integer resonance so we expect a large drop in DA below 100~MeV.

\begin{figure}[htbp]
  \centering
  \includegraphics[width=0.9\columnwidth]{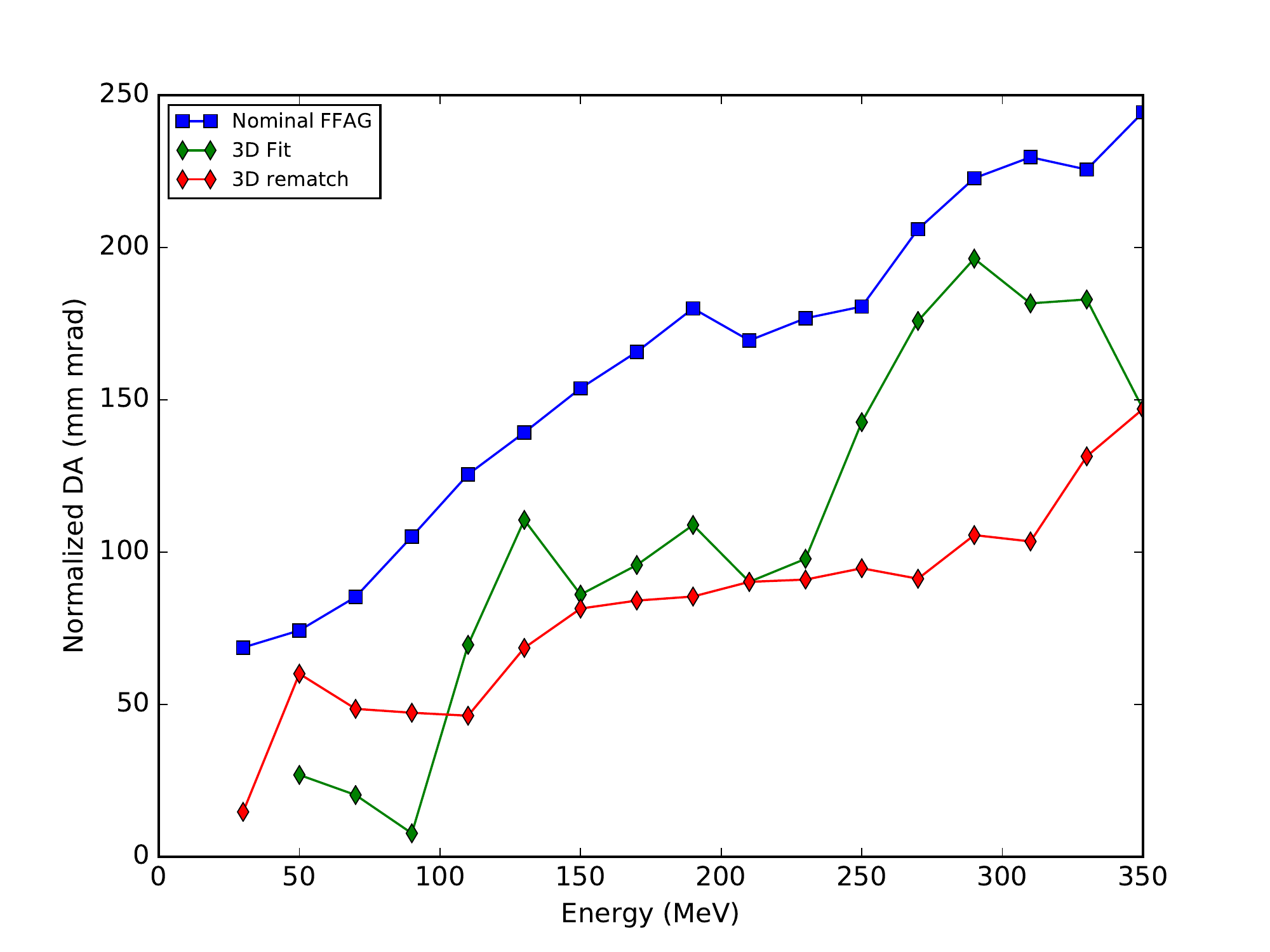}
  \caption{DA as a function of energy for 3D magnets and rematched fields.}
\label{fig:3d_norma_da_energy_DA}
\end{figure}

We simulate the 1000-turn DA in the 3D magnet using both the \texttt{POLARMESH} and fitted \texttt{DIPOLES} magnet elements. Figure~\ref{fig:3d_norma_da_energy_DA} shows a significant reduction compared to the nominal FFAG design, especially for energies below 100~MeV where the vertical tune crosses below the quarter-integer resonance. It is clear that the uncorrected map does not result in a sufficiently large stable region.

In order to improve the DA the magnets must be rematched so that the tune no longer crosses the harmful resonances. Ideally the tune can be flattened such that the dynamics are similar to the nominal FFAG design. This rematching is carried out by varying the multipole components of the fitted analytic field, and optimizing to obtain the nominal tune at all energies. The shape of the fringe falloff is held fixed as it is expected that a small change to the body field will not have a dramatic effect on the fringe, so that after another iteration of the magnet design the tune flatness will be retained.

We found that a rematch varying multipole coefficients up to decapole in the bodies of the magnets was sufficient to flatten the tune such that it no longer crossed harmful resonances.
The light blue lines (3D rematched) in Fig.~\ref{fig:3d_norma_da_energy_NU_Y} and~\ref{fig:3d_norma_da_energy_NU_Z} show the improvement in tune flatness when the magnets are rematched up to decapole order. Figure~\ref{fig:3d_norma_da_energy_DA} shows the DA for the rematched magnets in light blue. The rematched tune is sufficient to increase the DA to around 50~mm\,mrad in the critical region below 100~MeV and for it to remain above 50~mm\,mrad at higher energies.

\begin{figure}[htbp]
  \centering
  \includegraphics[width=0.9\columnwidth]{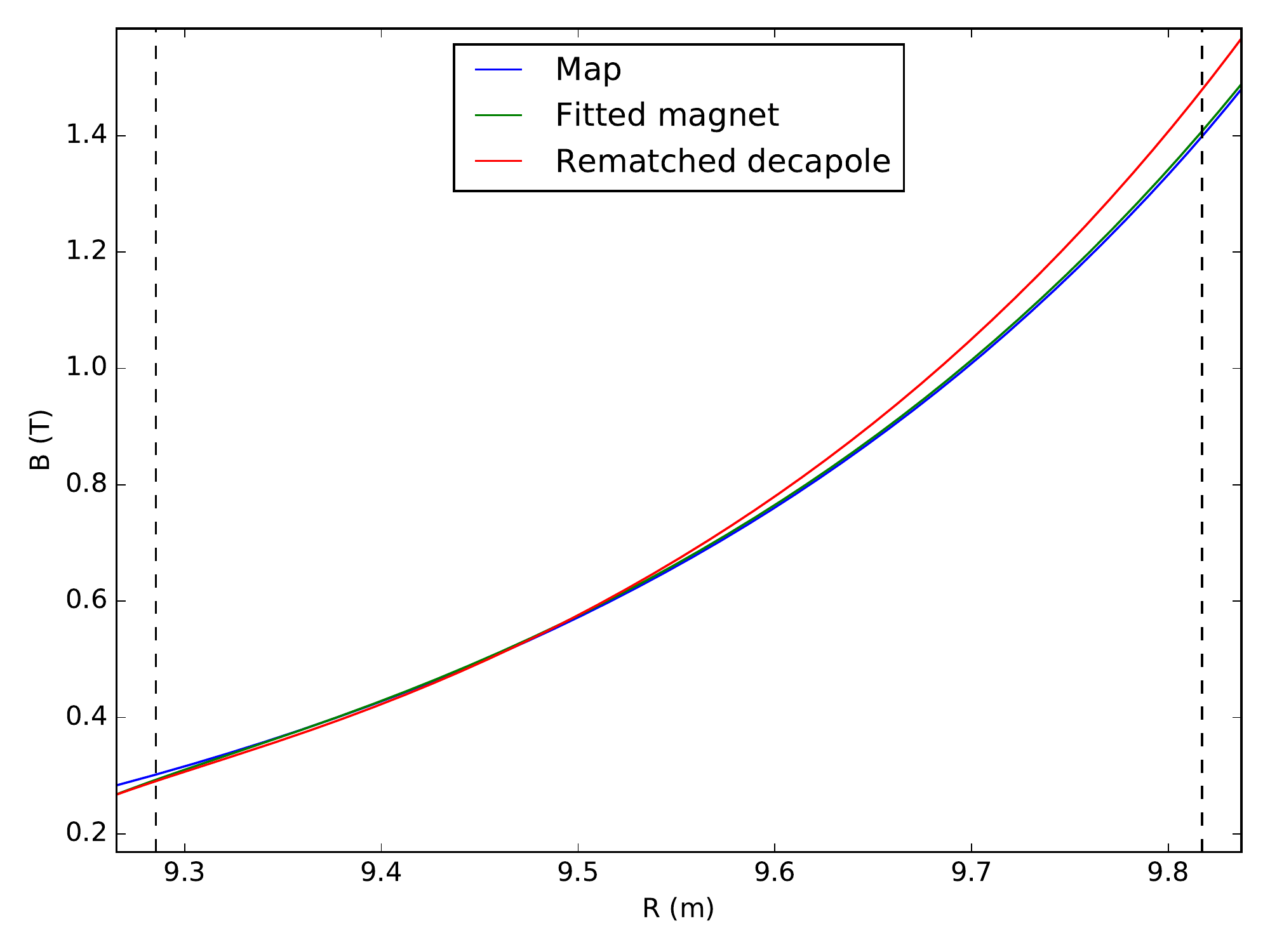}
  \caption{Field profile for rematched F magnet.}
\label{fig:rematched_profile_f}
\end{figure}

\begin{figure}[htbp]
  \centering
  \includegraphics[width=0.9\columnwidth]{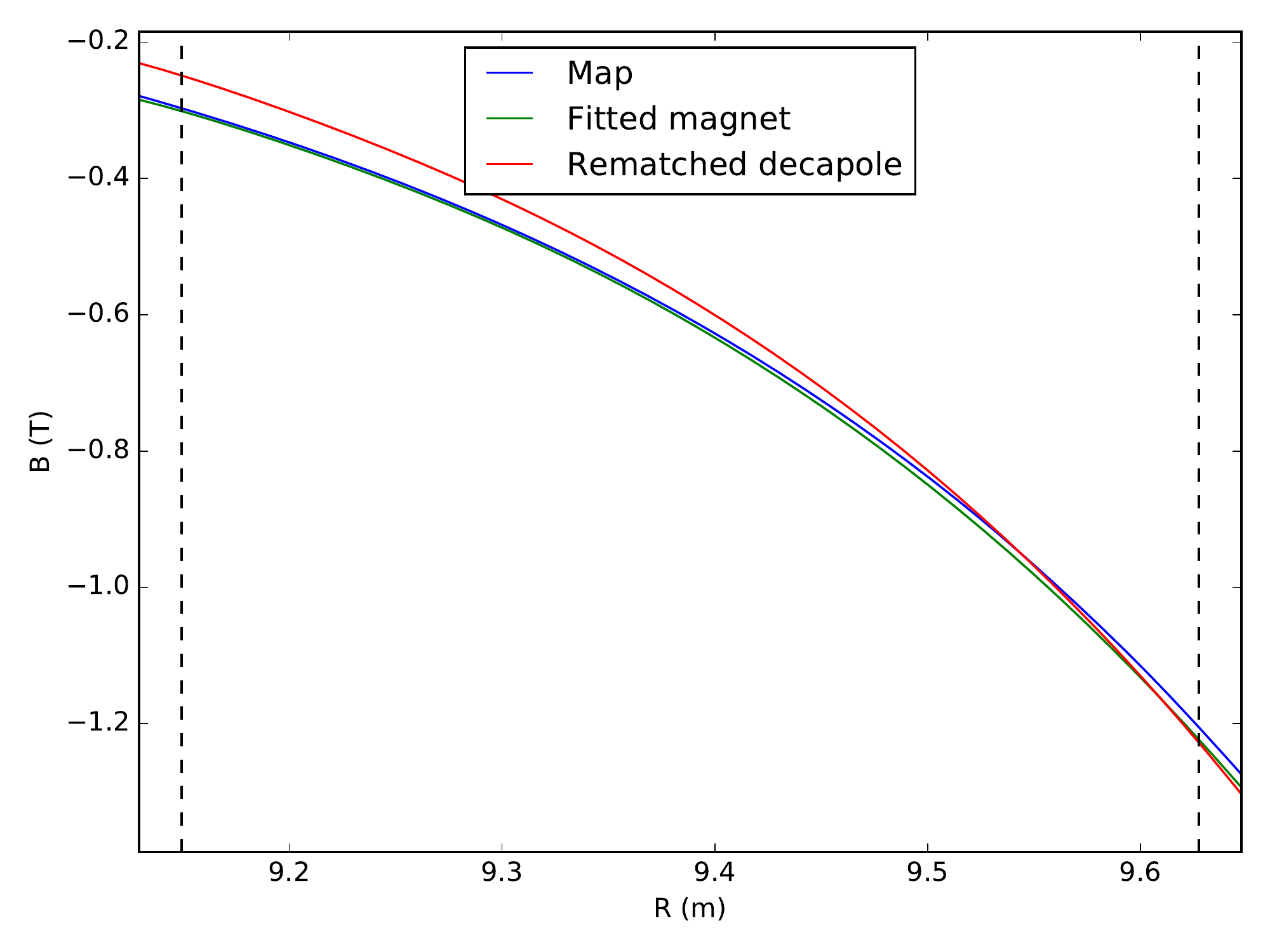}
  \caption{Field profile for rematched D magnet.}
\label{fig:rematched_profile_d}
\end{figure}

The main result from this work is the corrected (rematched) field profile for each magnet that recovers the original beam dynamics obtained with the idealistic (formula) fields. These field profiles together with the original (Map) field profiles obtained from the 3D Opera models are plotted in Figs.~\ref{fig:rematched_profile_f} and~\ref{fig:rematched_profile_d} for the F and D-magnets, respectively. The plots labelled ``Fitted'' are practically identical to the ``Map'' fields and represent an intermediate result needed to obtained the corrected (rematched) fields.
In the F magnet the difference between the fitted field and the rematched field is less than 5\%. In the D at small radius there is a 20\% reduction in the field, falling to 1\% at large radius. These modifications to the body field are sufficient to rematch the magnets that would otherwise give significantly altered dynamics to the original design.


\begin{figure}[htbp]
  \centering
  \includegraphics[width=0.9\columnwidth]{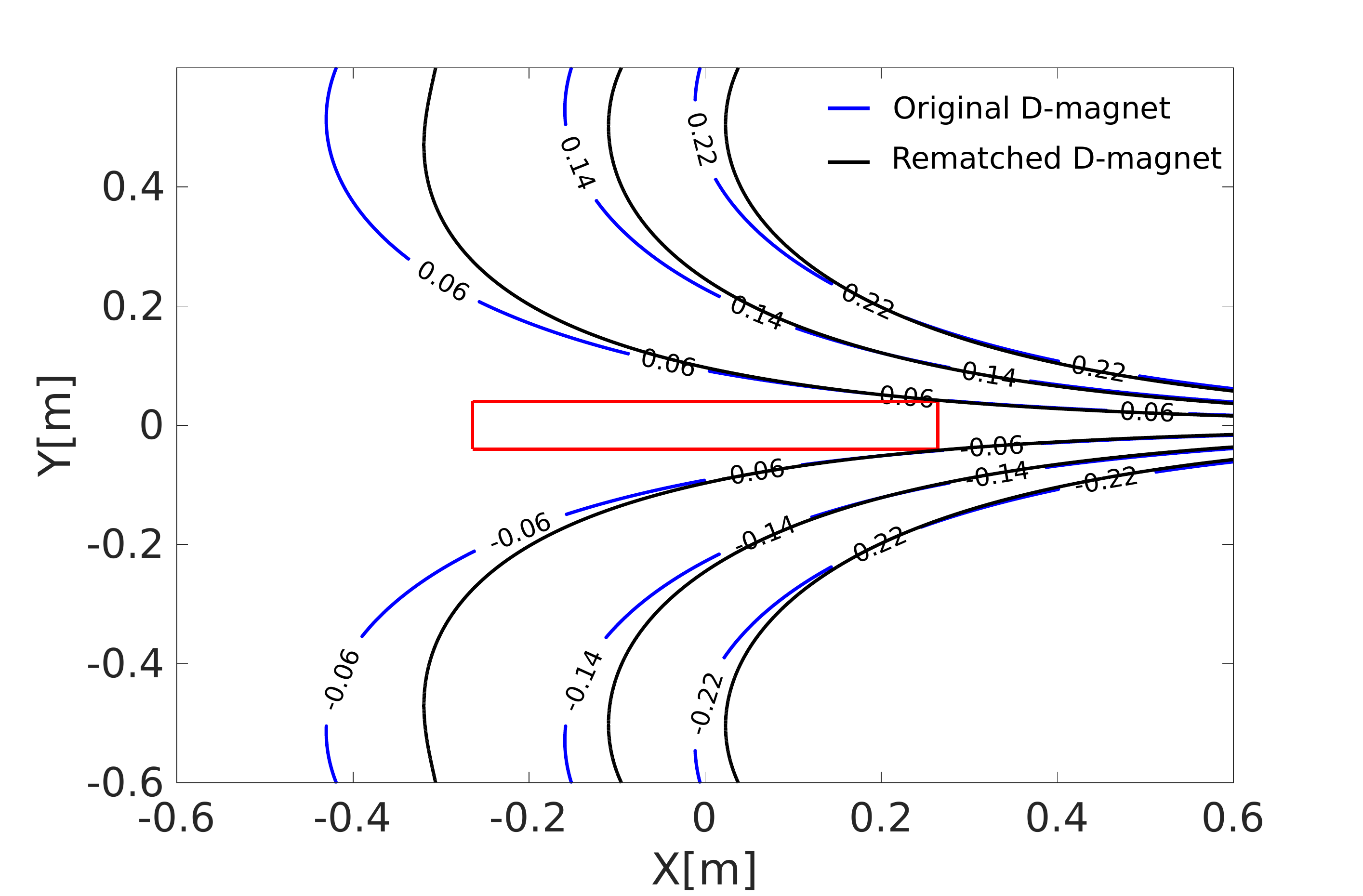}
  \caption{Lines of constant scalar potential for the original and adjusted D magnet.}
\label{fig:new_equipot_d}
\end{figure}

We now take the difference between the fitted and rematched field profiles, and add it to the original field profile given by Eq.~\ref{eq:scaling_law} in order to obtain the updated pole shapes of the two magnets that generate the rematched field profiles shown in Figs.~\ref{fig:rematched_profile_f} and~\ref{fig:rematched_profile_d}. The first step in this process is to obtain the updated lines of constant scalar potential. These are shown in Fig. \ref{fig:new_equipot_d} for the D-magnet (cf. Fig.~\ref{fig:fig1kiril}) and represent the zeroth-order approximation to the pole shape of the rematched D-magnet. At this stage the procedure described in Section \ref{sec:magnet_design_simulations} can be implemented to yield the revised magnet designs that generate the rematched field profiles followed by an analysis of the beam dynamics produced by these updated magnet designs. The process is repeated until the desired level of performance is reached. 

Table \ref{tab:final_params} shows the parameters of the F and D magnets achieved by our design process.

\begin{table}[!ht]
\caption{Magnet parameters.}
    \centering
    \begin{ruledtabular}
    \begin{tabular}{l r r}
	&Focusing	&Defocusing\\
\hline
Number of poles	&2	&2\\
$B_0$ (T)	&1.50	&-2.08\\
$r_0$ (m)	&9.82	&9.82\\
$k$	&27.50	&27.50\\
Max abs./rel. field error 2D	& $< 10^{-4}$	& $< 10^{-4}$ \\
Max abs./rel. field error 3D	& $< 10^{-2}$	& $< 10^{-2}$ \\
Horizontal aperture (m)	&0.53	&0.48\\
Approx. length (m)	&1.0	&1.0\\
Max abs. Field (T)	&1.40	&1.22\\
Min abs. Field (T)	&0.30	&0.30\\
    \end{tabular}
    \end{ruledtabular}
\label{tab:final_params}
\end{table}

\section{Conclusion}

The NORMA design for a medical proton accelerator has been previously demonstrated with idealized magnet modelling.
More detailed modelling using realistic magnet designs is important to show that the design is robust, and to understand the retuning needed to maintain the required dynamics and stability of the original design.
In this paper we have shown that the NORMA lattice can be modelled with magnet designs from 2D and 3D FEM simulations, taking the design from an idealized lattice to a detailed study with realistic magnets.

The F and D magnets were designed in 2D and 3D using OPERA. The field maps from the these models were imported into the PyZgoubi tracking code so that dynamics of the proton bunches could be compared to the idealized lattice.
For the 2D models - which give a radial profile - no significant effect on the dynamics was found. Random multipole errors, with similar-sized deviations as between the 2D profile and the analytic field, were found to have only a small effect on the stability of the lattice. 
Performing the magnet modelling and optimization work entirely in 3D is not optimal because of the complexity of the problem. As shown in the text, highly accurate 2D pole shapes for the two magnets are easy to obtain. These solutions allow a quick and realistic assessment of the available vertical and horizontal magnet apertures, the attainable good field region and, equivalently, the energy range of the machine to be made. In addition, the 2D pole solutions are a convenient starting point for the actual design work in 3D. The 3D models in turn provide a realistic representation of the actual magnetic fields and allow a detailed study and mitigation of field errors, fringe field and magnet cross-talk effects.

To see the importance of the fringe fields, simulations with altered fringe extents were carried out. It was found that changing the fringe length had a significant effect on the focusing of the lattice, and that for any given fringe extent the lattice must be retuned. By rematching the field strength and index in the magnets the working point could be re-optimized and sufficient DA recovered.
3D magnet models predict greater shifts from the nominal fields.
However, we show that it still possible to make small adjustments to the body fields that recover a sufficient DA over the energy range from injection to extraction; further iterations of the 3D design may be used to refine it.
More detailed modelling of the overlap region between the magnets is also needed, though it is expected that any changes to the dynamics can be accounted for by similar re-optimization.

This work demonstrates the importance of relying on rigorous 3D magnet simulations to model an FFAG accelerator, in order to obtain a realistic assessment of the overall machine performance. It also demonstrates methods of retuning an accelerator design to account for these effects, and to recover the original dynamic properties and beam stability.

\section{Acknowledgements}

We wish to thank Fran\c{c}ois M\'{e}ot for all his help with the Zgoubi code, and Jean-Baptiste Lagrange and Jaroslaw Pasternak for discussions and advice.

Research supported by STFC grant number ST/K002503/1 ``Racetrack FFAGs for medical, PRISM and energy applications''. The work undertaken in this paper was funded by the Cockcroft Institute Core Grant (STFC grant reference ST/G008248/1).

\end{document}